\documentclass[twocolumn]{aastex631} 

\usepackage[version=3]{mhchem} 
\usepackage{seqsplit}
\usepackage{enumitem}
\usepackage{tabularx}
\usepackage{graphicx}%
\usepackage{multirow}%
\usepackage{booktabs} 

\shorttitle{Chemical mapping of temperate sub-Neptune atmospheres}
\shortauthors{Yang \& Hu}

\begin{document}

\title{Chemical mapping of temperate sub-Neptune atmospheres:\\ Constraining the deep-interior \ce{H2O}/\ce{H2} ratio from the atmospheric \ce{CO2}/\ce{CH4} ratio}

\correspondingauthor{Jeehyun Yang; Renyu Hu}
\email{jeehyun.yang@jpl.nasa.gov; renyu.hu@jpl.nasa.gov}

\author[0000-0002-1551-2610]{Jeehyun Yang}
\affiliation{Jet Propulsion Laboratory, California Institute of Technology,
Pasadena, CA 91109, USA}

\author[0000-0003-2215-8485]{Renyu Hu}
\affiliation{Jet Propulsion Laboratory, California Institute of Technology,
Pasadena, CA 91109, USA}
\affiliation{Division of Geological and Planetary Sciences, California Institute of Technology, Pasadena, CA 91125, USA}


\begin{abstract}

Understanding the planetary envelope composition of sub-Neptune-type exoplanets is challenging due to the inherent degeneracy in their interior composition scenarios.  Particularly, the planetary envelope's \ce{H2O}/\ce{H2} ratio, which can also be expressed as the O/H ratio, provides crucial insights into its original location relative to the ice line during planetary formation. Using self-consistent radiative transfer modeling and a rate-based automatic chemical network generator combined with 1D photochemical kinetic-transport atmospheric modeling, we investigate various atmospheric scenarios of temperate sub-Neptunes, ranging from \ce{H2}-dominated to \ce{H2O}-dominated atmospheres with equilibrium temperatures (\textit{T}\textsubscript{eq}) of 250–400 K. This study includes examples such as K2-18 b (\textit{T}\textsubscript{eq} = 255 K), LP 791-18 c (\textit{T}\textsubscript{eq} = 324 K), and TOI-270 d (\textit{T}\textsubscript{eq} = 354 K). Our models indicate that the atmospheric \ce{CO2}/\ce{CH4} ratio can be used to infer the deep-interior \ce{H2O}/\ce{H2} ratio. Applying this method to recent JWST observations, our findings suggest K2-18~b likely has an interior that is 50\% highly enriched in water, exceeding the water content in a 100$\times$\textit{Z}\textsubscript{$\odot$} scenario and suggesting a planetary formation mechanism involving substantial accretion of ices. In contrast, our model suggests that approximately 25\% of TOI-270~d's interior is composed of \ce{H2O}, which aligns with the conventional metallicity framework with a metallicity higher than 100$\times$\textit{Z}\textsubscript{$\odot$}. Furthermore, our models identify carbonyl sulfide (OCS) and sulfur dioxide (\ce{SO2}) as strong indicators for temperate sub-Neptunes with at least 10\% of their interior composed of water. These results provide a method to delineate the internal composition and formation mechanisms of temperate sub-Neptunes ($T_{\rm eq}<\sim$ 500 K) via atmospheric characterization through transmission spectroscopy.
\end{abstract}


\section{Introduction} \label{sec:intro}
Sub-Neptune-sized planets separated from super-Earth-sized planets by the radius valley near 1.8 \textit{R}${\oplus}$ generally have a volatile-rich envelope and are the most common types of planets detected so far \citep{fulton_california-kepler_2017, Fulton_2018, Van_Eylen_2018, Hardegree-Ullman_2019}. Despite their prevalence, our understanding of the interior composition of sub-Neptunes is very limited due to the inherent degeneracy in their interior structure composition. The radius and mass of sub-Neptunes can be explained by either (i) a massive, high-molecular-weight (e.g., \ce{H2O}) volatile layer or (ii) a lighter \ce{H2}/\ce{He}-dominated envelope on top of a rock/iron core \citep{Rogers_2010, Valencia_2010, Luque_2022}.

For this reason, precise characterization of the internal composition and resulting chemistry of sub-Neptunes has become increasingly important from a formation and evolution perspective. This characterization might help distinguish between different pathways of sub-Neptune formation. For example, `\ce{H2O}-rich' sub-Neptunes are believed to form farther from their star, beyond the ice line, where they can efficiently accrete large amounts of water and volatiles in the form of solid material before migrating to their current close-in orbits \citep{Lambrechts-2014, Morbidelli-2015, Venturini_2020}. In contrast, `\ce{H2}-rich' sub-Neptunes with lower volatile content are thought to originate within the ice line, where volatiles are not present as ice, and most of their mass is accumulated through the accretion of drifting rocky pebbles \citep{Johansen-2017}. Therefore, determining the deep-interior \ce{H2O}/\ce{H2} ratio of sub-Neptunes spanning the radius valley is essential for enhancing our understanding of planetary formation and evolution.

Temperate sub-Neptunes (200 K $\leq$\textit{T}\textsubscript{eq} $\leq$ 400 K) are excellent targets for studying planetary interior compositions using JWST observations. Cold exoplanets could potentially have liquid water oceans (note that Earth has \textit{T}\textsubscript{eq} $\sim$ 255 K, and has ocean), leading to atmospheric compositions primarily regulated by interactions with liquid water. As a result, atmospheric composition does not directly represent the envelope composition. (e.g., gaseous ammonia is highly soluble in liquid water, which leads to depletion in the upper atmosphere). On the other hand, many hotter exoplanets are predicted to contain \ce{CO} and \ce{CO2} in their atmospheres due to their higher thermochemical stability compared to \ce{CH4} at high temperatures, and thus these gases do not exclusively represent the deep-interior composition \citep{Moses_2011, Line_2011, Venot_2012, Moses_2016, Tsai_2018, Fortney_2020}. Additionally, studies have shown that exoplanets with \textit{T}\textsubscript{eq} between 500-800 K exhibit the most attenuated spectral features, likely due to thick clouds and hazes \citep{Morley_2015, Brande_2024}. 

Therefore, temperate sub-Neptuens with \textit{T}\textsubscript{eq} lower than 400 K are favorable observation targets. However, it is important to note that water may condense below the photosphere (the region probed by JWST) in the cooler end of the temperate range (i.e., \textit{T}\textsubscript{eq} $\sim$ 250 K), complicating the direct inference of internal \ce{H2O} content from atmospheric observations. Recent advancements in observational techniques with high enough signal-to-noise ratios have made sub-Neptunes favorable targets as well. Numerous JWST observations of various sub-Neptunes are already available such as K2-18 b \citep{Madhusudhan-2023} and TOI-270 d \citep{Benneke-2024_jwst}, or will be available soon (LP 791-18 c). Since high-altitude clouds or hazes that form at \textit{T}\textsubscript{eq} $\leq$ 400 K are unlikely to obscure the transmission spectra \citep{Morley_2015}, JWST observations can provide detailed and valuable information about the molecular species present in these atmospheres.

Several efforts have been made to infer the elemental composition of the deep atmospheres of sub-Neptunes using both theoretical and observational studies \citep{Thorngren_2016, Thorngren_2019, Luque_2022, Burn-2024, Benneke-2024_jwst}. A previous study has modeled the atmospheres of water-rich sub-Neptunes (from solar abundance to 100\% \ce{H2O}) using 1D radiative-convective equilibrium modeling coupled with their associated transmission and thermal emission spectra \citep{Kempton_2023}. However, this approach lacks disequilibrium chemistry and thus may miss infrared absorbers beyond those abundant in thermochemical equilibrium. To our knowledge, no attempts have been made to infer the envelope composition, particularly the \ce{H2O}/\ce{H2} ratio of potential water-rich temperate sub-Neptunes, by fully utilizing 1D photochemical modeling. \ce{CO2} is a carbon-bearing species indicative of high temperatures and oxidative atmospheres, while \ce{CH4} indicates low temperatures and reducing atmospheres \citep{Moses_2011, Moses_2016}. Consequently, the ratio between these two molecules can provide insights into the bulk envelope \ce{H2O}/\ce{H2} ratio of temperate sub-Neptunes, potentially helping us understand their original locations relative to the ice line within the protoplanetary disk \citep{Johansen-2017, Burn-2024}.

In this work, we employ self-consistent radiative transfer modeling and a state-of-the-art rate-based automatic chemical network generator combined with 1D photochemical kinetic-transport atmospheric modeling to investigate various atmospheric scenarios, ranging from \ce{H2}-dominated (representing gas accretion) to \ce{H2O}-dominated (representing ice accretion) atmospheres of temperate sub-Neptune-type exoplanets. We introduce a new method for inferring the deep-interior (or bulk envelope) \ce{H2O}/\ce{H2} ratio from the atmospheric \ce{CO2}/\ce{CH4} ratio and a new framework to classify temperate sub-Neptunes using this O/H ratio, different from the conventional multipliers of solar metallicity framework (i.e., \textit{n}$\times Z_{\odot}$). This new variable (i.e., O/H ratio) not only guides the interpretation of JWST observations of sub-Neptunes but also provides new insights into their original location within the protoplanetary disk during planetary formation.

\section{Methods} \label{sec:methods}
\subsection{elemental parameterization of planetary envelope accretion}\label{subsec:elements}
To investigate the diverse envelope accretion scenarios of sub-Neptunes, we established seven distinct \ce{H2O}/\ce{H2} accretion scenarios, represented by corresponding oxygen-to-hydrogen (O/H) ratio scenarios, ranging from \ce{H2}-rich to \ce{H2O}-rich envelopes. This variability was numerically achieved by adjusting the O/H ratio beyond the framework of conventional solar metallicity, $Z_{\odot}$ \citep{Lodders-2020}. As detailed in Table A\ref{tab:atom_abundances} in Appendix \ref{sec:appendix_a}, the third column lists standard multipliers of solar metallicity ($Z_{\odot}$) at 1$\times$, 10$\times$, 100$\times$, 1000$\times$, and 10000$\times$ (denoted as \ce{H2}:\ce{H2O} accretion ratio =100:0 for simplicity). Based on this and while retaining the original abundances of carbon, nitrogen, and sulfur (i.e., each elemental abundance follows its solar elemental abundance multiplied by an integer, \textit{n}), we systematically varied the O to H ratio according to the equations detailed in Appendix \ref{sec:appendix_a}. From now on, maintaining C, N, and S at each of their solar elemental abundances multiplied by \textit{n} is denoted as [C+N+S]=\textit{n}$\times Z_{\odot}$. 

This spectrum of accretion scenarios (see the description of $x_{\text{acc}}$ in Appendix \ref{sec:appendix_a}) reflects the variations in planetary formation locations, particularly concerning the ice line, as depicted in Figure A\ref{fig:elemental_abundances} in Appendix \ref{sec:appendix_a}. Overall, we analyzed 74 planetary atmospheric structures (\textit{T--P} profiles) and 80 planetary atmospheric photochemical models, the details of which are described in Appendix \ref{sec:appendix_a}.

\subsection{The temperature-pressure (T--P) profiles for various planetary envelope accretion scenarios}\label{sec:TP_profiles}
For each of the elemental parameterizations of planetary envelope accretion mentioned in Section \ref{subsec:elements}, we calculated 74 \textit{T--P} profiles under radiative-convective equilibrium using the climate module of the ExoPlanet Atmospheric Chemistry \& Radiative Interaction Simulator (EPACRIS-Climate, Scheucher et al., \textit{in prep}), and the results are presented in Figure B\ref{fig:TP-profiles} in Appendix \ref{sec:appendix_b} (1000$\times$ and 10000$\times$ solar metallicity cases for K2-18~b are omitted for simplicity). Further detail on this radiative-convective equilibrium modeling can be found in Appendix \ref{sec:appendix_c}.

\subsection{Automoatic chemical reaction network generation for \ce{H2O}-rich chemistry}\label{sec:rmg}

A detailed chemical reaction network for \ce{H2O}-rich atmospheres was constructed using the Reaction Mechanism Generator (RMG) \citep{Gao_2016, RMG-database}, a Python-based open-source software. RMG generates chemical networks using a rate-based iteration algorithm and has been extensively described in previous literature \cite{Gao_2016, rmg-v3}. This method's application to exoplanet atmospheric studies is detailed in \cite{Yang_2024}, and further detail on this chemical reaction network generation for \ce{H2O}-rich chemistry can be found in Appendix \ref{sec:appendix_c}.

The final photochemical network comprised 92 species (56 originally available in the EPACRIS species, and 36 species newly generated by RMG, not available in the EPACRIS library, as listed in Table C\ref{tbl: newspecies} in Appendix \ref{sec:appendix_c}). The network included 2009 reactions (343 original EPACRIS reactions = 40 photochemistry reactions + 248 bi-molecular reactions + 34 ter-molecular reactions + 21 thermo-dissociation reactions, and 1666 reactions newly generated by RMG. Except for the 40 photochemistry reactions, the other 1969 reactions are forward-reverse reaction pairs. This comprehensive network was used in 1-D photochemical kinetic-transport atmospheric modeling of various temperate sub-Neptune atmospheres, as described in Section \ref{sec:epacris}.

\subsection{1-D photochemical kinetic-transport atmospheric modelings}\label{sec:epacris}
Based on the elemental scenarios outlined in Section \ref{subsec:elements}, the \textit{T--P} profiles calculated in Section \ref{sec:TP_profiles}, and the chemical network tailored for water-rich systems in Section \ref{sec:rmg}, we performed 1D photochemical kinetic-transport atmospheric modeling of the 80 scenarios using the chemistry module of EPACRIS \citep{Yang_2024}. This modeling was conducted to simulate the steady-state vertical mixing ratios of chemical species in various atmospheric scenarios of temperate sub-Neptunes. Further details on the eddy diffusion coefficients and the stellar fluxes used in this study can be found in Appendix \ref{sec:appendix_d}.

After the models had converged and reached the steady state, we computed the synthetic transmission spectra of K2-18~b and TOI-270~d based on the molecular mixing ratio profiles, using the transmission spectra generation module of EPACRIS \citep{Hu_2013}, and compared the resulting transmission spectra with published JWST observations \citep{Madhusudhan-2023, Benneke-2024_jwst}.

\section{Results and Discussions} \label{sec:Results_and_Discussions}
\subsection{Constraining the bulk envelope \ce{H2O}/\ce{H2} ratio} \label{sec: constraining_deep_interior_H2O_H2_ratio}

\begin{figure*}
    \centering
    \includegraphics[width=1\textwidth]{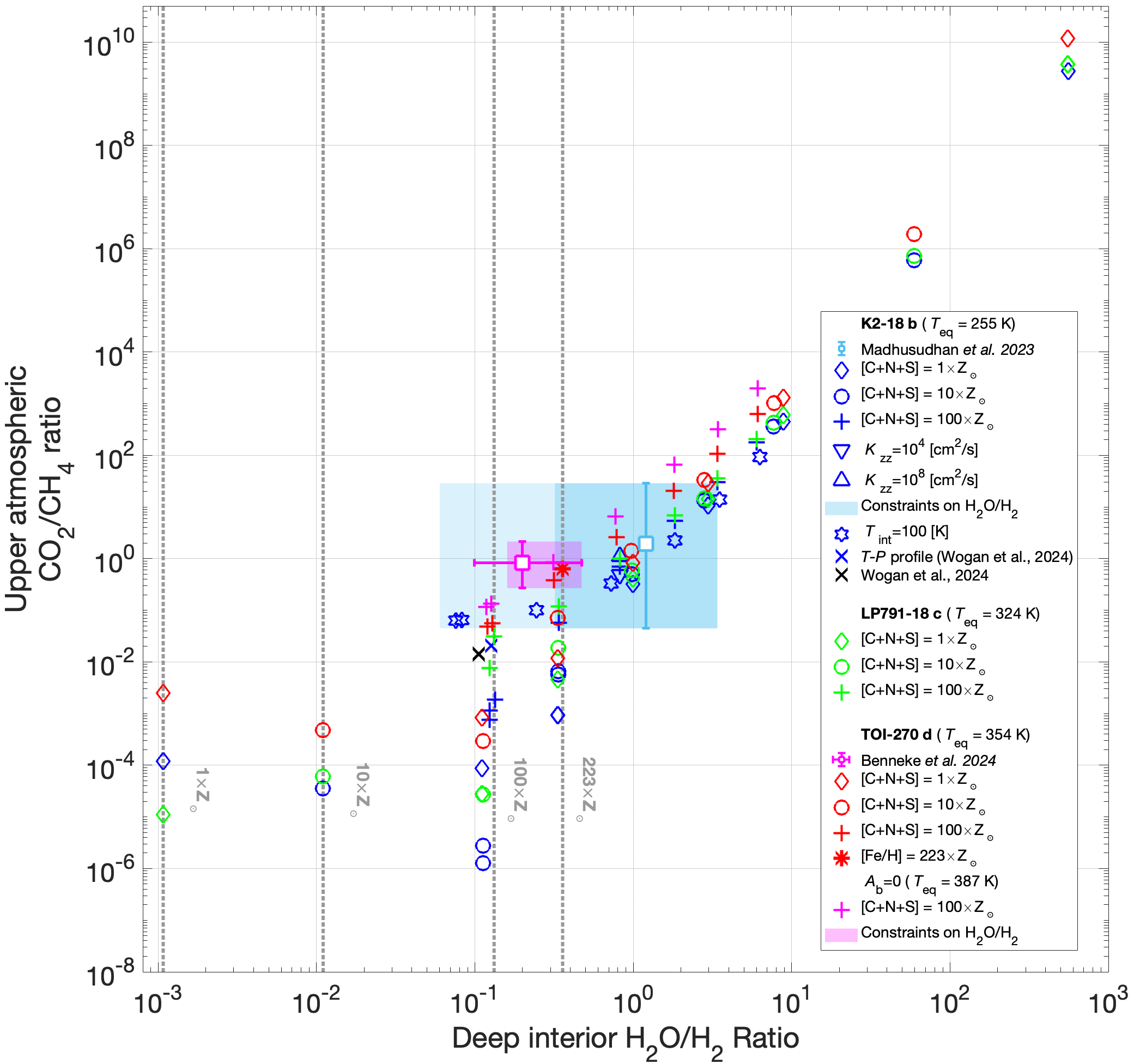}
    \caption{\footnotesize The figure shows the upper atmospheric (\textit{P}=0.1--2 mbar, corresponding to the JWST's primary probing range; \cite{Rustamkulov-2023}) \ce{CO2}-to-\ce{CH4} ratios plotted against the deep interior (\textit{P}=20 bar) \ce{H2O}-to-\ce{H2} ratios. Unless otherwise noted, the planetary equilibrium temperature (i.e., \textit{T}\textsubscript{eq}) is assumed based on \textit{A$_b$}=0.3 and an intrinsic temperature (\textit{T}\textsubscript{int}) of 60 K, with $Z_{\odot}$ denoting solar metallicity. The symbols in the figure represent different elemental parameterizations for the planetary envelope accretion scenarios. Diamonds indicate seven O/H scenarios (i.e., \ce{H2}:\ce{H2O} from 100:0 to 0:100) with elemental compositions of carbon, nitrogen, and sulfur ([C+N+S]) of 1$\times Z_{\odot}$. Circles represent the same O/H ratios but with [C+N+S] of 10$\times Z_{\odot}$, while crosses correspond to 100$\times Z_{\odot}$. Upward- and downward-pointing triangles indicate the scenario with an \ce{H2}:\ce{H2O} ratio of 50:50 and [C+N+S] of 100$\times Z_{\odot}$, but assuming eddy diffusion coefficients of 10$^8$ and 10$^4$ [cm$^2\cdot$s$^{-1}$], respectively. Hexagrams also illustrate these O/H ratios with [C+N+S] of 100$\times Z_{\odot}$, but assuming intrinsic temperature (\textit{T}\textsubscript{int}) of 100 K. X symbols represent the K2-18~b scenario with 100$\times Z_{\odot}$, based on the \textit{T--P} profile from \cite{wogan2024jwst}. Blue symbols indicate the model simulated by EPACRIS (in this work), and black symbols represent the model by \cite{wogan2024jwst}. The primary differences between the two models lie in the chemical networks and the eddy diffusion coefficient profiles used for modeling. An asterisk indicates 223$\times Z_{\odot}$. A white-colored square with light blue error bars indicates the JWST observation of K2-18~b \citep{Madhusudhan-2023}, and a white-colored square with magenta error bars indicates the JWST observation of TOI-270~d \citep{Benneke-2024_jwst}. Each color represents the model simulations for corresponding planets: blue and black for K2-18~b, green for LP791-18~c, red for TOI-270~d, magenta for TOI-270~d assuming \textit{A$_b$}=0. The color-shaded box regions indicate the implied constraints on the deep-interior \ce{H2O}-to-\ce{H2} ratios--or envelope O/H ratio--of corresponding planets based on current studies (light blue for K2-18~b and magenta for TOI-270~d). A much lighter blue-shaded box region indicates the constraints when accounting for hotter intrinsic temperature (\textit{T}\textsubscript{int}=100 K) case of K2-18~b with 100$\times Z_{\odot}$. Grey dotted lines represent standard multipliers of solar metallicity (1, 10, 100, and 223$\times$), aiding in contextualizing the current study against conventional benchmarks in solar metallicity.}
    \label{fig:CO2_CH4_ratios}
\end{figure*}

\subsubsection{Using the \ce{CO2}/\ce{CH4} ratio as a constraint on the \ce{H2O}/\ce{H2} ratio} \label{sec: co2toch4-ratio_as_a_constraint}
Figure \ref{fig:CO2_CH4_ratios} displays the upper atmospheric \ce{CO2}/\ce{CH4} ratios plotted against the deep interior \ce{H2O}/\ce{H2} ratios for various planetary envelope accretion scenarios, as detailed in Section \ref{subsec:elements}. As an example, for the case of \ce{H2}:\ce{H2O} = 50:50 and [C+N+S] = 100$\times$ solar metallicity scenario of K2-18~b, as illustrated in the left panel of Figure \ref{fig:vmr}, we determined the \ce{CO2} (red) and \ce{CH4} (green) mixing ratios within the pressure range of 0.1--2 mbar, located within the grey-shaded area, to derive the upper atmospheric \ce{CO2}/\ce{CH4} ratios. Similarly, we obtained the \ce{H2O} (blue) and \ce{H2} (black) mixing ratios at a pressure of 200 bar (i.e., $2\times10^5$ mbar) at the bottom to derive the deep-interior \ce{H2O}/\ce{H2} ratios. This approach allowed us to explore the relationship between the upper atmospheric \ce{CO2}/\ce{CH4} ratios and the deep-interior \ce{H2O}/\ce{H2} ratios across various planetary envelope accretion scenarios. By analyzing these upper atmospheric \ce{CO2}/\ce{CH4} ratios and comparing them to the \ce{CO2}/\ce{CH4} ratios retrieved from the JWST observation of K2-18~b \citep{Madhusudhan-2023} and TOI-270~d \citep{Benneke-2024_jwst}, we can further constrain the deep-interior \ce{H2O}/\ce{H2} ratio of each planet, respectively. 

In general, as expected, we can see a clear consistent linear pattern across all planetary envelope accretion scenarios with more than 25\% water (corresponding to 223$\times Z_{\odot}$, shown by the grey dotted-line in Figure \ref{fig:CO2_CH4_ratios}). This is because a \ce{H2O}-rich envelope creates an oxidative atmosphere favorable for \ce{CO2} formation over \ce{CH4}, thus increasing the \ce{CO2}/\ce{CH4} ratio as the \ce{H2O}-content inside the planetary envelope increases. By leveraging this trend, we can effectively constrain the deep-interior \ce{H2O}/\ce{H2} ratio using the upper atmospheric \ce{CO2}/\ce{CH4} ratio.

First, in the case of K2-18~b, as depicted in Figure \ref{fig:CO2_CH4_ratios}, assuming \textit{A}$_b$=0.3 (i.e., \textit{T}\textsubscript{eq}=255 K), an intrinsic temperature (i.e., \textit{T}\textsubscript{int}) of 60 K, and an eddy diffusion coefficient of $10^6$ [cm$^2$/s], the thicker blue-shaded area represents the deep-interior \ce{H2O}/\ce{H2} constraints for K2-18~b, which ranges from $\sim$25 to $\sim$90\% water-rich scenarios. This indicates that K2-18~b's envelope contains at least $\sim$25\% water. This statement contrasts with the observations by \cite{Madhusudhan-2023}, which did not detect significant contributions from \ce{H2O} and thus did not provide any atmospheric retrievals on O/H ratios. As discussed later in Section \ref{sec:K2-18b}, in the case of K2-18~b, its low equilibrium temperature of 255 K causes water to condense out at \textit{P}$\sim10^2$ mbar, making water detection challenging for JWST observation (see blue line in the left panel of Figure \ref{fig:vmr} and the upper panel of Figure \ref{fig:transmission_spectra}). This underscores the importance of accurately addressing water condensation in atmospheric modeling to extract hidden information.

Adjusting for an \textit{T}\textsubscript{int}=100 K, to account for the sensitivity of the intrinsic temperature to the upper atmospheric \ce{CO2}/\ce{CH4} ratios due to changes in deep-interior \textit{T--P} structure, the lighter blue-shaded area now also marginally includes the conventional 100$\times$solar metallicity scenario, as denoted as blue hexagrams in Figure \ref{fig:CO2_CH4_ratios}. This behavior is primarily attributed to elevated temperatures in the deep interior of K2-18~b, which favor the formation of \ce{CO2} over \ce{CH4} \citep{Moses_2011}, thus elevating the upper atmospheric \ce{CO2}/\ce{CH4} ratios in \ce{H2}-rich cases (e.g., 100$\times$solar metallicity) to align with the lower-end of the JWST observational constraints on \ce{CO2}/\ce{CH4} ratio. 

This elevated \ce{CO2}/\ce{CH4} ratio in \ce{H2}-rich envelope is also consistent with the modeling result of \cite{wogan2024jwst}, which utilized a \textit{T--P} profile that is $\sim$150 K higher at 200 bar compared to the \textit{T--P} profile computed by EPACRIS assuming the same \textit{T}\textsubscript{int} of 60 K. However, it is 100 K lower compared to the \textit{T--P} profile computed by EPACRIS assuming the \textit{T}\textsubscript{int} of 100 K, which encompasses the uncertainty. Although the \ce{CO2}/\ce{CH4} ratio predicted by \cite{wogan2024jwst} is slightly below the lower end of the JWST observational constraints on \ce{CO2}/\ce{CH4} ratio (see black X symbol in Figure \ref{fig:CO2_CH4_ratios}), the synthesized transmission spectrum contained both \ce{CO2} and \ce{CH4} features. To determine whether the elevated \ce{CO2}/\ce{CH4} ratios in the upper atmosphere of the \ce{H2}-rich case are predominantly due to the \textit{T--P} profile rather than differences in the chemical networks used, we conducted photochemical modeling of the K2-18~b atmosphere. We used the same old 100$\times$ solar metallicity from \cite{Lodders-2009}, where C/O = 0.46, compared to the recently updated value of C/O = 0.55 from \cite{Lodders-2020}. We also applied the same \textit{T--P} profile as in \cite{wogan2024jwst}. The resulting \ce{CO2}/\ce{CH4} ratio is denoted by a blue X symbol in Figure \ref{fig:CO2_CH4_ratios} and exhibits almost identical value to that modeled by \cite{wogan2024jwst} (see black X symbol). This emphasizes that accurate prediction of the interior \textit{T--P} structure is crucial for using the upper atmospheric \ce{CO2}/\ce{CH4} ratio to infer the deep-interior \ce{H2O}/\ce{H2} ratios. 

We also tested the sensitivity of the eddy diffusion coefficients to the upper atmospheric \ce{CO2}/\ce{CH4} ratios for the \ce{H2}:\ce{H2O}=50:50 and [C+N+S]=100$\times$solar metallicity scenario of K2-18~b, denoted by upward- (assuming $10^8$ [cm$^2$/s]) and downward-pointing blue triangles ($10^4$ [cm$^2$/s]) in Figure \ref{fig:CO2_CH4_ratios}, which shows relatively less sensitivity compared to the sensitivity driven by varying intrinsic temperatures.

In the case of TOI-270~d, as depicted in Figure \ref{fig:CO2_CH4_ratios}, assuming \textit{A}$_b$=0 (i.e., \textit{T}\textsubscript{eq}=387 K), an intrinsic temperature (i.e., \textit{T}\textsubscript{int}) of 60 K, and an eddy diffusion coefficient of $10^6$ [cm$^2$/s], the magenta-shaded area represents the deep-interior \ce{H2O}/\ce{H2} constraints for TOI-270~d, which indicates that TOI-270~d's envelope contains approximately 15--25\% water. These constraints align well with the quenched-chemistry atmospheric retrievals on O/H ratios (i.e., horizontal error bars in Figure \ref{fig:CO2_CH4_ratios}) of the deep atmospheres (1--10 bar) based on the JWST observations \citep{Benneke-2024_jwst}. Since \cite{Benneke-2024_jwst} also provides a constraint on the TOI-270~d's metallicity of approximately $223\times Z_{\odot}$ (note that this corresponds to \ce{H2}:\ce{H2O}=80:20), we ran additional photochemical modeling assuming \textit{A}$_b$=0.3 (i.e., \textit{T}\textsubscript{eq}=354 K), an intrinsic temperature (i.e., \textit{T}\textsubscript{int}) of 60 K, and an eddy diffusion coefficient of $10^6$ [cm$^2$/s]. As shown in Figure \ref{fig:CO2_CH4_ratios}, the resulting \ce{CO2}/\ce{CH4} ratio, denoted by a red asterisk, is consistent with the JWST measured \ce{CH4} and \ce{CO2}, and also falls within the deep-interior \ce{H2O}/\ce{H2} ratio constraint presented in this work. This demonstrates the robustness of the current framework for inferring the deep-interior O/H ratio using 1-D photochemical modeling, particularly when combined with observational data.

\begin{figure*}
    \centering
    \includegraphics[width=0.95\textwidth]{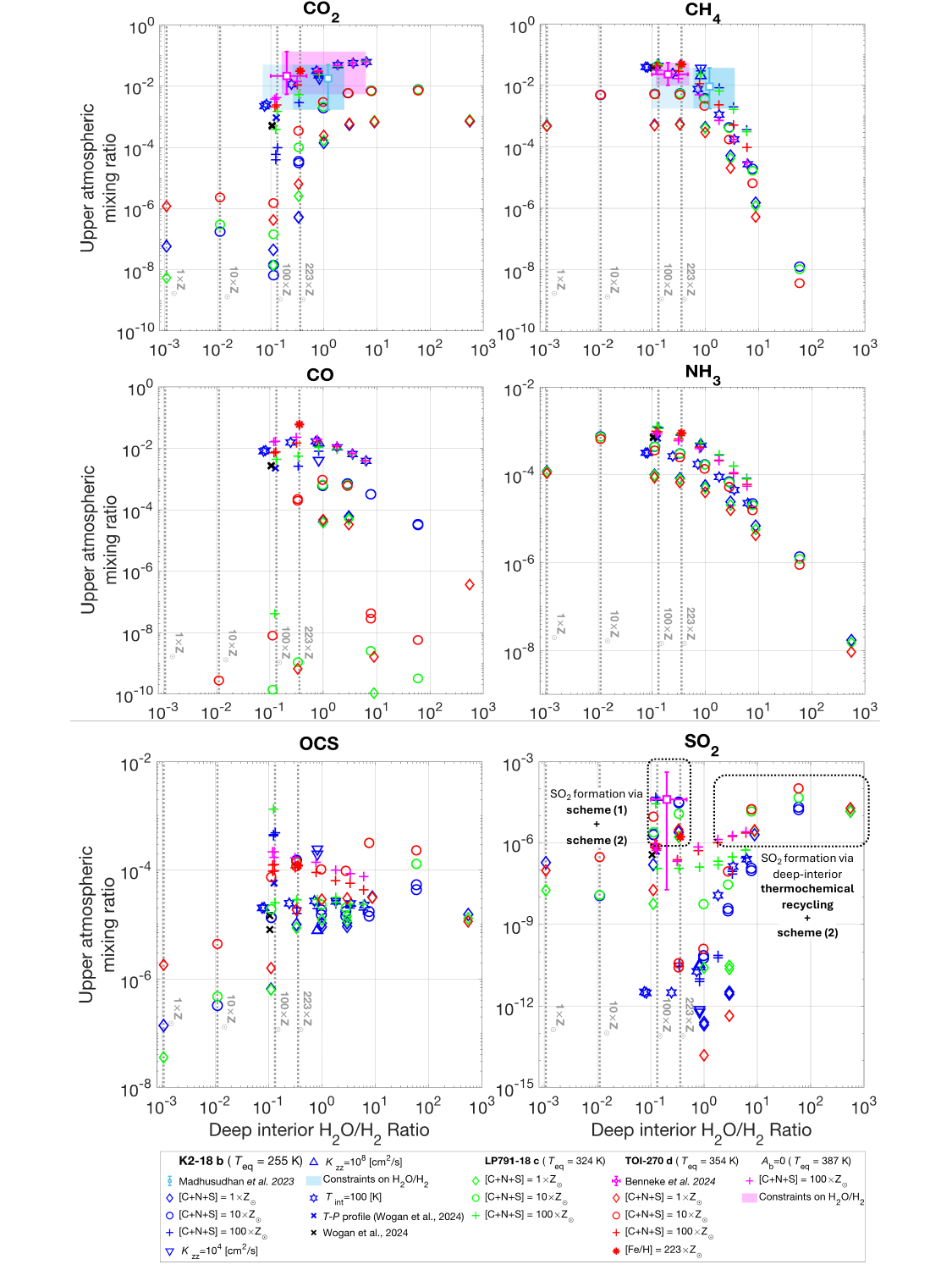}
    \caption{\footnotesize The upper atmospheric (\textit{P}=0.1--2mbar) \ce{CO2}, \ce{CH4}, \ce{CO}, \ce{NH3}, \ce{OCS}, and \ce{SO2} molecular mixing ratios plotted against the deep interior (\textit{P}=20 bar) \ce{H2O}-to-\ce{H2} ratios, respectively. All notations follow Figure \ref{fig:CO2_CH4_ratios}. The black dotted box indicates the major formation pathways that contribute to \ce{SO2} formation, respectively.}
    \label{fig:all_species}
\end{figure*}

\subsubsection{Using individual species abundances as further constraints} \label{sec: species_mixing_ratio_as_a_constraint}

As described in Section \ref{sec: co2toch4-ratio_as_a_constraint}, the upper atmospheric \ce{CO2}/\ce{CH4} ratio can be a powerful tool for constraining the bulk envelope \ce{H2O}/\ce{H2} ratio. However, this metric does not retain information about the absolute abundances of carbon, nitrogen, and sulfur species in the envelope. It is therefore crucial to investigate whether the abundance of individual gases could provide additional constraints for characterizing the atmosphere and envelope of temperate sub-Neptunes. For instance, could measurements of additional gases help distinguish \textit{T}\textsubscript{int} of 60 K versus 100 K for a planet like K2-18~b? In this section, we explore the behavior of several individual species across various planetary envelope scenarios of temperate sub-Neptunes.

Figure \ref{fig:all_species} shows the upper atmospheric \ce{CO2}, \ce{CH4}, \ce{CO}, \ce{NH3}, \ce{OCS}, and \ce{SO2} molecular mixing ratios plotted against the deep interior \ce{H2O}/\ce{H2} ratios for various planetary envelope scenarios. Generally, a clear pattern emerges between \ce{CO2} and \ce{CH4}. These two molecules exhibit opposite behaviors--\ce{CO2} predominates in water-rich envelopes (i.e., deep interior \ce{H2O}/\ce{H2}$\geq$0.1), while \ce{CH4} is more abundant in \ce{H2}-rich envelopes. Notably, in scenarios with a deep interior \ce{H2O}/\ce{H2}$\sim$1, the presence of both gases is nearly balanced.

Similar to \ce{CO2}, carbon monoxide (CO) is favored in an oxidizing environment. However, if there is an excess of oxidizers compared to CO, it will remain fully oxidized as carbon monoxide. Therefore, it is challenging to exclusively determine whether the atmosphere is \ce{H2}-dominated or \ce{H2O}-dominated based solely on the CO abundance. Nevertheless, our model indicates that the CO mixing ratio consistently exceeds 1000 ppm if the C-abundance is greater than 100$\times Z_{\odot}$, as shown in Figure \ref{sec: species_mixing_ratio_as_a_constraint}, identifying CO as a strong indicator of a C-rich planetary envelope. Although the retrieved detection significance of CO abundance is relatively less credible compared to \ce{CO2} and \ce{CH4} due to its sparse infrared absorption line-list density, the largest upper limits of $10^{-3}$ retrieved from \cite{Madhusudhan-2023} suggest that the carbon abundance should be less than 100$\times Z_{\odot}$. Consequently, \cite{wogan2024jwst}'s modeling that assumes 100$\times Z_{\odot}$ for K2-18~b (black X symbols in Figure \ref{sec: species_mixing_ratio_as_a_constraint}) might not be able to explain the JWST observations of K2-18~b. However, our model, which assumes a carbon abundance of 10$\times Z_{\odot}$ and a \ce{H2}:\ce{H2O}=50:50 ratio can still explain all retrieved JWST constraints on \ce{CO2}, \ce{CH4}, and \ce{CO} \citep{Madhusudhan-2023}. In the case of TOI-270~d, the retrieved upper limit of CO abundance was $10^{-1.46}\sim0.035$, consistent with all scenarios of our model.

Similar to \ce{CH4}, \ce{NH3} indicates reducing envelope conditions (\ce{H2}-rich) as shown in Figure \ref{fig:all_species}, and could be a sensitive indicator for constraining the deep-interior \ce{H2O}/\ce{H2} ratio. For instance, if adopting \ce{NH3}'s largest upper limits of $10^{-4.46}\sim0.000035$ retrieved for K2-18~b from \cite{Madhusudhan-2023}, all scenarios with nitrogen abundance of 100$\times Z_{\odot}$ are already ruled out. However, it is important to note that the entire envelope could be intrinsically nitrogen-poor, as N-bearing species can effectively dissolve into the mantle beneath the envelope \citep{Shorttle_2024}. This explanation was also proposed for \ce{NH3}-depletion (upper limits of $10^{-4.27}$) observed in TOI-270~d \citep{Benneke-2024_jwst}.

The upper atmospheric mixing ratios of carbonyl sulfide (OCS) with respect to the deep interior \ce{H2O}-to-\ce{H2} ratios exhibit very interesting features. As shown in Figure \ref{fig:all_species}, regardless of the exact OCS abundance being sensitive to \textit{K}\textsubscript{zz} (see blue upward- and downward-pointing triangles), the upper atmospheric OCS mixing ratio is generally predicted to be above $\sim10^{-5}$ (sufficient to appear as a specific spectral feature at 4.8--4.9 $\mu$m, as shown in Figure \ref{fig:transmission_spectra}) for deep-interiors enriched with more than 10\% \ce{H2O} across various temperate sub-Neptune scenarios, with \textit{T}\textsubscript{eq} ranging from 255 K to 387 K. \cite{Moses_2013} investigated compositional diversity of carbonyl sulfide in hot Neptune-sized exoplanets using a thermochemical-equilibrium model in the framework of solar metallicity, but its abundance was less than 1 ppm even at 500$\times Z_{\odot}$ (see Figure 5 in \cite{Moses_2013}). However, in our modeling, the OCS formation mechanism is different since we are investigating the case of temperate sub-Neptune exoplanets being a \ce{H2O}-rich interior. As described in detail later in Section \ref{subsec:K2-18b_chemistry} and \ref{subsec:TOI-270d_chemistry}, the OCS formation was attributed to the enriched water content in the deep interior, resulting in an oxidizing envelope that favors carbon-bearing species in the form of \ce{CO} and \ce{CO2}. Sulfur, primarily from \ce{H2S}, then reacts with this CO to form OCS (Scheme \ref{eqn:OCS_deep_interior_formation}), which is then transported to the upper atmospheric region observable by the JWST. Additional OCS formation occurs at higher altitudes due to increased sulfur abundance from continuous \ce{H2S} photodissociation in the presence of CO (Scheme \ref{eqn:OCS_photochemical_formation}). The OCS formation is thus a result of the oxidative chemistry due to the water-rich envelope. This suggests that carbonyl sulfide (OCS) is a strong indicator of a water-rich envelope for temperate sub-Neptunes, highlighting OCS as an indicator of temperate sub-Neptune envelopes with over 10\% \ce{H2O}.

As shown in Figure \ref{fig:all_species}, species such as \ce{CO2}, \ce{CH4}, \ce{NH3}, and \ce{OCS} exhibit intuitive and consistent behavior with respect to the deep-interior O/H ratio. However, the upper atmospheric \ce{SO2} mixing ratio displays a somewhat indirect pattern compared to other species, indicating its formation pathways are varied. Multiple factors--including \textit{T--P} structure, absolute sulfur abundance, and O/H ratio--affect \ce{SO2} formation, making it more complex. Nonetheless, \ce{SO2} formation pathways can be divided into two representative regimes, as detailed in Appendix \ref{sec:appendix_e}.

Our models spanning various water-rich envelope scenarios don't always predict \ce{SO2} levels above 10 ppm (which is sufficient to appear as spectral features in JWST observations). However, any \ce{SO2} level above 10 ppm in our model indicates a deep-interior with more than 10\% \ce{H2O} inside. For this reason, along with OCS, our model suggests that any future detection of \ce{SO2} in temperate sub-Neptune atmospheres indicates a water-rich interior of at least more than 10\% \ce{H2O}.

Although we only used \ce{CO2}+\ce{CH4} in the current study due to the limited or absent observational constraints on N- or S-bearing species, other species such as \ce{NH3}, \ce{OCS}, or \ce{SO2} could serve as additional tools to more precisely constrain the deep-interior \ce{H2O}/\ce{H2} ratio along with the absolute N- or S-abundance as described in this section.

\subsubsection{Constraining the building blocks of temperate sub-Neptunes} \label{subsec: constraining_building_blocks}

\begin{figure*}
    \centering
    \includegraphics[width=1\textwidth]{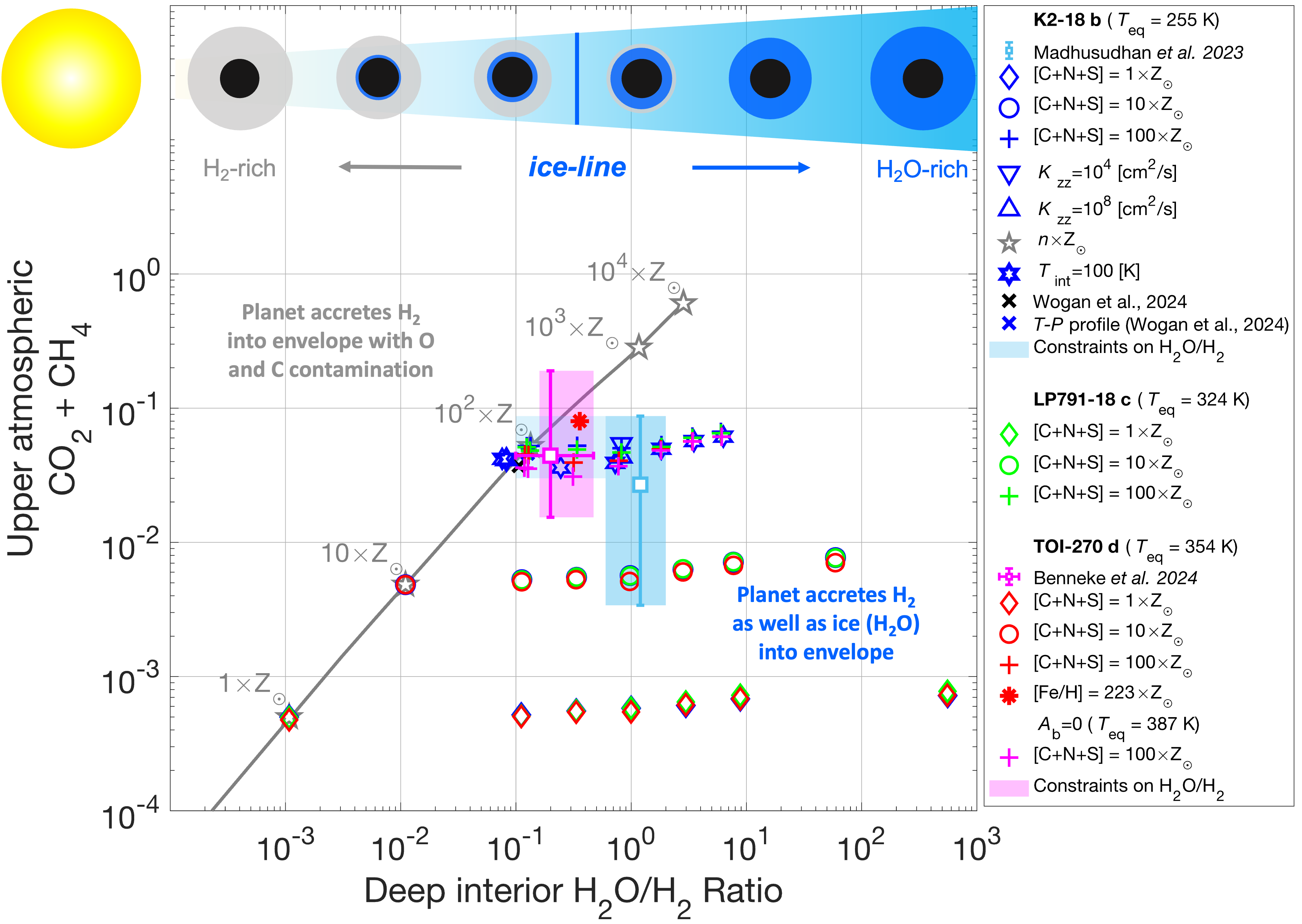}
    \caption{\footnotesize The figure shows a schematic diagram of the feeding zone of planets in the protoplanetary disk with respect to the ice-line on the top and the upper atmospheric (\textit{P}=0.1--2 mbar) \ce{CO2}+\ce{CH4} mixing ratios plotted against the deep interior (\textit{P}=20 bar) \ce{H2O}-to-\ce{H2} ratios on the bottom. It should be noted that the position of the ice-line is depicted arbitrarily to better illustrate the concept of the current study. All notations follow Figure \ref{fig:CO2_CH4_ratios}. The only difference is that the constraints on the \ce{H2O}/\ce{H2} ratio for each planet are now based on the combination of the \ce{CO2}/\ce{CH4} ratio (Section \ref{sec: co2toch4-ratio_as_a_constraint}) and individual abundances (Section \ref{sec: species_mixing_ratio_as_a_constraint}).}
    \label{fig:total_C}
\end{figure*}

Figure \ref{fig:total_C} presents a comprehensive plot after integrating all the constraints from the \ce{CO2}/\ce{CH4} ratio and each of the mixing ratios of \ce{CO2} and \ce{CH4}, as described in Section \ref{sec: co2toch4-ratio_as_a_constraint} and \ref{sec: species_mixing_ratio_as_a_constraint}. Notably in this Figure \ref{fig:total_C}, we can constrain the carbon abundance included in the planetary envelope to within the range of $10-100\times Z_{\odot}$ for K2-18~b, and approximately $40-500\times Z_{\odot}$ for TOI-270~d using absolute \ce{CO2} and \ce{CH4} mixing ratio. Particularly, the carbon abundance for K2-18~b constrained within the range of $10-100\times Z_{\odot}$ aligns well with the carbon abundance implied from the CO abundance shown in Section \ref{sec: species_mixing_ratio_as_a_constraint} and Figure \ref{fig:all_species}. 

According to Figure \ref{fig:total_C}, the most probable scenario for TOI-270~d is close to the \ce{H2}:\ce{H2O}=75:25 and [C+N+S]=100$\times Z_{\odot}$ assuming either \textit{A}$_b$=0 (magenta cross) or 0.3 (red cross). Although the uncertainty is substantial, the \ce{SO2} mixing ratio constrained from the JWST observation of TOI-270~d by \cite{Benneke-2024_jwst} also falls within this O/H ratio range as shown in the \ce{SO2} panel of Figure \ref{sec: species_mixing_ratio_as_a_constraint}. In the case of K2-18~b, two scenarios--\ce{H2}:\ce{H2O}=50:50 and \ce{H2}:\ce{H2O}=25:75, both with [C+N+S]=100$\times Z_{\odot}$ scenarios (represented by blue crosses) fall within the constraints from the chemical mapping (i.e., thick blue shaded-area). 

Furthermore, Figure \ref{fig:total_C} reveals some new insights: TOI-270~d can still fit within the conventional solar metallicity framework, depicted by a grey solid line (particularly simulated for various K2-18~b planetary envelope scenarios) in Figure \ref{fig:total_C}. If we maintain this framework, TOI-270~d would require a metallicity significantly above 100$\times Z_{\odot}$ value. This is corroborated by a red asterisk marker on the plot, representing the retrieved constraint on the metallicity of $\sim$223$\times Z_{\odot}$ from the JWST observation by \cite{Benneke-2024_jwst}. However, K2-18~b cannot be explained within this conventional solar metallicity framework (note: a C/O ratio of 0.134--0.138 is required to explain the JWST observations of K2-18~b shown in Figure \ref{fig:total_C}). Instead, its characteristics (i.e., thicker blue-shaded region) suggest efficient ice accretion within the protoplanetary disk, implying that planetary formation occurred close to or beyond the ice line \citep{Burn-2024}. This underscores the importance of the O/H ratio as a pivotal variable in classifying exoplanets, potentially revealing their original locations of planetary formation. By combining this chemical mapping method with many upcoming observational data, we can determine whether the planet accretes \ce{H2} into the envelope with other elements, or efficiently accretes \ce{H2O} in the form of ice into its envelope, as shown in Figure \ref{fig:total_C}. 

\subsection{The atmosphere of K2-18~b}\label{sec:K2-18b}
\subsubsection{Overall chemistry in the atmosphere of K2-18~b}\label{subsec:K2-18b_chemistry}
\begin{figure*}
    \centering
    \includegraphics[width=1\textwidth]{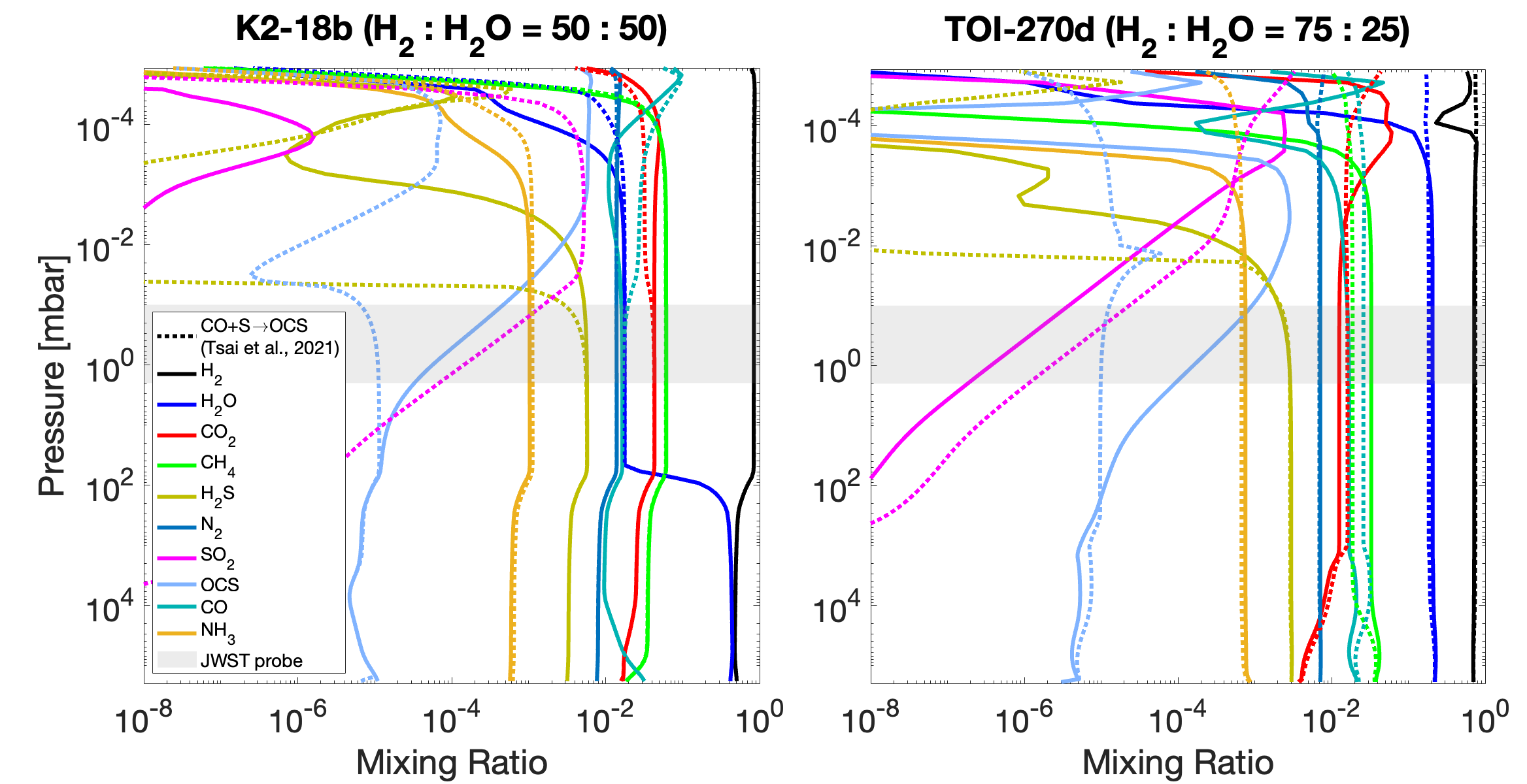}
    \caption{\footnotesize Vertical molecular mixing ratio profiles of major species for the constrained scenarios for K2-18~b and TOI-270~d in Section \ref{sec: constraining_deep_interior_H2O_H2_ratio}: K2-18~b with an \ce{H2}:\ce{H2O} ratio of 50:50 and [C+N+S] of 100$\times Z_{\odot}$ assuming \textit{A}\textsubscript{b} = 0.3 (left), and TOI-270~d with an \ce{H2}:\ce{H2O} ratio of 75:25 and [C+N+S] of 100$\times Z_{\odot}$ assuming \textit{A}\textsubscript{b} = 0 (right). Each color indicates the corresponding species: \ce{H2} in black, \ce{H2O} in blue, \ce{CO2} in red, \ce{CH4} in green, \ce{H2S} in dark yellow, \ce{N2} in dark blue, \ce{SO2} in magenta, \ce{OCS} in light blue, \ce{CO} in teal, and \ce{NH3} in light brown. The dotted lines indicate major species' vertical mixing ratio profiles when using the rate-coefficient of Scheme \ref{eqn:OCS_recombination} adopted from \cite{Tsai_2021}. The grey-shaded area indicates the JWST primary probing range (i.e., \textit{P}=0.1--2 mbar; \cite{Rustamkulov-2023})}
    \label{fig:vmr}
\end{figure*}

The left panel of Figure \ref{fig:vmr} shows the simulated vertical molecular mixing ratio profiles of major species for the scenario close to the most probable scenario for K2-18~b (i.e., \ce{H2}:\ce{H2O}=50:50 with [C+N+S]=100$\times Z_{\odot}$) according to the chemical mapping detailed in Section \ref{sec: constraining_deep_interior_H2O_H2_ratio}. The first notable thing to look at is the condensation behavior of water at \textit{P}$\sim10^2$ mbar. This is consistent with the computed \textit{T--P} profiles for various scenarios of K2-18~b as shown in Figure B\ref{fig:TP-profiles} in Appendix \ref{sec:appendix_b}. As illustrated in Figure B\ref{fig:TP-profiles}, water condensation would begin around \textit{P}$\sim10^2$ mbar, where the condensation curves for \ce{H2O}\citep{Buck-1981} intersect with the \textit{T--P} profiles for various O/H ratios with [C+N+S]=100$\times Z_{\odot}$ scenarios for K2-18~b (see blue solid lines in Figure B\ref{fig:TP-profiles}). Consequently, the JWST observable region of the upper atmosphere will have a \ce{H2}-dominated atmosphere with less than $\sim$1\% of \ce{H2O}, enhancing spectral features due to a decreased mean molecular weight \citep{Miller-Ricci_2009}. The mixing ratios of other molecular species such as \ce{H2}, \ce{CH4}, \ce{CO2}, \ce{N2}, \ce{H2S}, and \ce{NH3} in the JWST probe region are primarily determined by the deep-interior thermal chemistry and are quenched at pressures greater than $10^4$ mbar, before being transported upward. This quenching behavior is expected since the equilibrium temperatures of temperate sub-Neptunes are low, leading to longer chemical lifetimes relative to the vertical mixing timescale. At higher altitudes (i.e., pressures lower than 0.1 mbar), UV-driven photochemistry dominates, photodissociating species including \ce{H2O}, \ce{CO2}, \ce{H2S}, and \ce{NH3}.

Another notable feature is the significant presence of carbonyl sulfide (OCS) in the JWST observable region, as shown in the left panel of Figure \ref{fig:vmr}. OCS primarily forms in the deep interior through the reaction between sulfur from \ce{H2S} and carbon monoxide, which is the favorable carbon-bearing species in \ce{H2O}-rich oxidizing envelopes: 
\begin{equation}
    \begin{split}
        \ce{H2S}&\rightarrow\ce{H2}+\ce{S}\\
        \ce{S}+\ce{CO}&\xrightarrow{\text{M}}\ce{OCS}\\
    \end{split}
    \label{eqn:OCS_deep_interior_formation}
\end{equation}
with M representing any third-body molecule. Further details of sulfur chemistry in the atmosphere of K2-18~b can be found in Appendix \ref{sec:appendix_f}. It is noteworthy that the rate coefficient of the termolecular reaction:
\begin{equation}
    \ce{S}+\ce{CO}\xrightarrow{\text{M}}\ce{OCS}
    \label{eqn:OCS_recombination}
\end{equation}
has significant uncertainty, as previously discussed in \cite{Ranjan_2020, Tsai_2021}. Potentially, this uncertainty can lead to a significant difference in the composition of dominant sulfur-bearing species (e.g., OCS as dominant or \ce{SO2} as dominant) in the upper atmosphere, where temperatures are usually very low, below 500 K. This is because even a small uncertainty in the activation energy in the exponential term of the Arrhenius rate equation can lead to substantial deviations from the actual reaction rate at these lower temperatures. Additionally, unlike fuel chemistry involving CHO species (or maybe including N), we lack extensive details on sulfur chemistry (e.g., sulfur chemistry induced by excited S(\textsuperscript{1}D) which would be important in the upper atmosphere, but poorly understood). For this reason, we need a systematic approach to address this issue in future studies.

In the current study, RMG adopts the OCS thermal dissociation rate coefficient measured by the shock tube experiments conducted by \cite{Oya-1994, Woiki-1995}. The rate coefficient for the reverse direction (i.e., Scheme \ref{eqn:OCS_recombination}) was then calculated using the Gibbs free energies of the corresponding species, which is the same rate coefficient adopted for reaction \ref{eqn:OCS_recombination} in \cite{Zahnle_2016}. The sensitivity of the vertical mixing ratio profiles of both K2-18~b and TOI-270~d's most probable scenarios to the OCS recombination rate is shown in Figure \ref{fig:vmr}, with the corresponding synthesized transmission spectra available in Appendix \ref{sec:appendix_f}. As illustrated in Figure \ref{fig:vmr}, although most major species show no significant changes in their predicted vertical mixing ratios, the dominant sulfur-bearing species now shifts from \ce{OCS} (solid lines which adopt the rate coefficient of Scheme \ref{eqn:OCS_recombination} adopted from \cite{Zahnle_2016}) to \ce{SO2} (dotted lines which adopt the rate coefficient of Scheme \ref{eqn:OCS_recombination} estimated from \cite{Tsai_2021}). Although the statement connecting any potential detection of \ce{SO2} or OCS in temperate sub-Neptune atmospheres to a water-enriched interior remains unchanged, this strongly suggests the need for future studies to accurately estimate the rate coefficient for Scheme \ref{eqn:OCS_recombination} through either quantum chemical calculations or experimental reaction kinetic measurements.


\subsubsection{Theoretical Transmission Spectra of the Atmosphere of K2-18~b generated by EPACRIS}\label{subsec:K2-18b_transmission_spectra}
The top panel of Figure \ref{fig:transmission_spectra} compares EPACRIS-generated theoretical transmission spectra with JWST observations of K2-18~b \citep{Madhusudhan-2023}. The EPACRIS prediction aligns well with the JWST data, particularly in capturing the \ce{CO2} (red) and \ce{CH4} (green) features identified by \cite{Madhusudhan-2023}. Although \ce{NH3} was not detected in the retrieval by \cite{Madhusudhan-2023}, EPACRIS-generated transmission spectra suggest its presence at around 3$\mu$m, as well as CO around 4.6$\mu$m, and notably, OCS around 4.8$\mu$m. This highlights the need for more detailed observations at these spectral windows to detect these molecules, which may have been overlooked due to low resolution or insufficient transits. Any potential detection of \ce{NH3} would be particularly significant as it can rule out the possibility of the Hycean scenario for K2-18~b \citep{Madhusudhan-2021, Yu-2021, Hu_2021b, Tsai_2021}, underscoring the importance of detailed observations. Additionally, the detection of OCS would be noteworthy since this molecule has not been detected in exoplanetary atmospheres so far.

\subsection{The atmosphere of TOI-270~d}\label{sec:TOI-270d}
\subsubsection{Overall chemistry in the atmosphere of TOI-270~d}\label{subsec:TOI-270d_chemistry}
The right panel of Figure \ref{fig:vmr} shows the simulated vertical molecular mixing ratio profiles of major species for the scenario close to the most probable scenario for TOI-270~d (i.e., \ce{H2}:\ce{H2O}=75:25 with [C+N+S]=100$\times Z_{\odot}$ assuming \textit{A}\textsubscript{b}=0) according to the chemical mapping detailed in Section \ref{sec: constraining_deep_interior_H2O_H2_ratio}. Different from the K2-18~b scenario, no significant condensation of water is observed due to the much higher equilibrium temperature of TOI-270~d (\textit{T}\textsubscript{eq}= 387 K) compared to that of K2-18~b  (\textit{T}\textsubscript{eq}= 255 K). As illustrated in Figure B\ref{fig:TP-profiles}, all \textit{T--P} profiles of various planetary envelope composition scenarios of TOI-270~d (see red and magenta lines in Figure B\ref{fig:TP-profiles}) lie above the water condensation line \citep{Buck-1981}. Consequently, the entire atmosphere contains a significant amount of water ($\geq$10\%), leading to decreased spectral features due to an increased mean molecular weight from \ce{H2O}. The overall chemistry regarding molecular species such as \ce{H2}, \ce{CH4}, \ce{CO2}, \ce{N2}, \ce{H2S}, \ce{NH3}, and \ce{OCS} is similar to that of K2-18~b, as described in Section \ref{subsec:K2-18b_chemistry}. In contrast, the upper atmosphere of TOI-270~d is richer in \ce{H2O} compared to K2-18~b's upper atmosphere where water condensation occurs. This leads to increased \ce{SO2} formation due to more \ce{H2O} being photolyzed by UV radiation, forming H and OH radicals that gradually oxidize reduced sulfur species (e.g., \ce{H2S}, S, and \ce{S2}) into \ce{SO2} \citep{Tsai_2023}.

\subsubsection{Theoretical Transmission Spectra of the Atmosphere of TOI-270~d generated by EPACRIS}\label{subsec:TOI-270d_transmission_spectra}
\begin{figure*}
    \centering
    \includegraphics[width=1\textwidth]{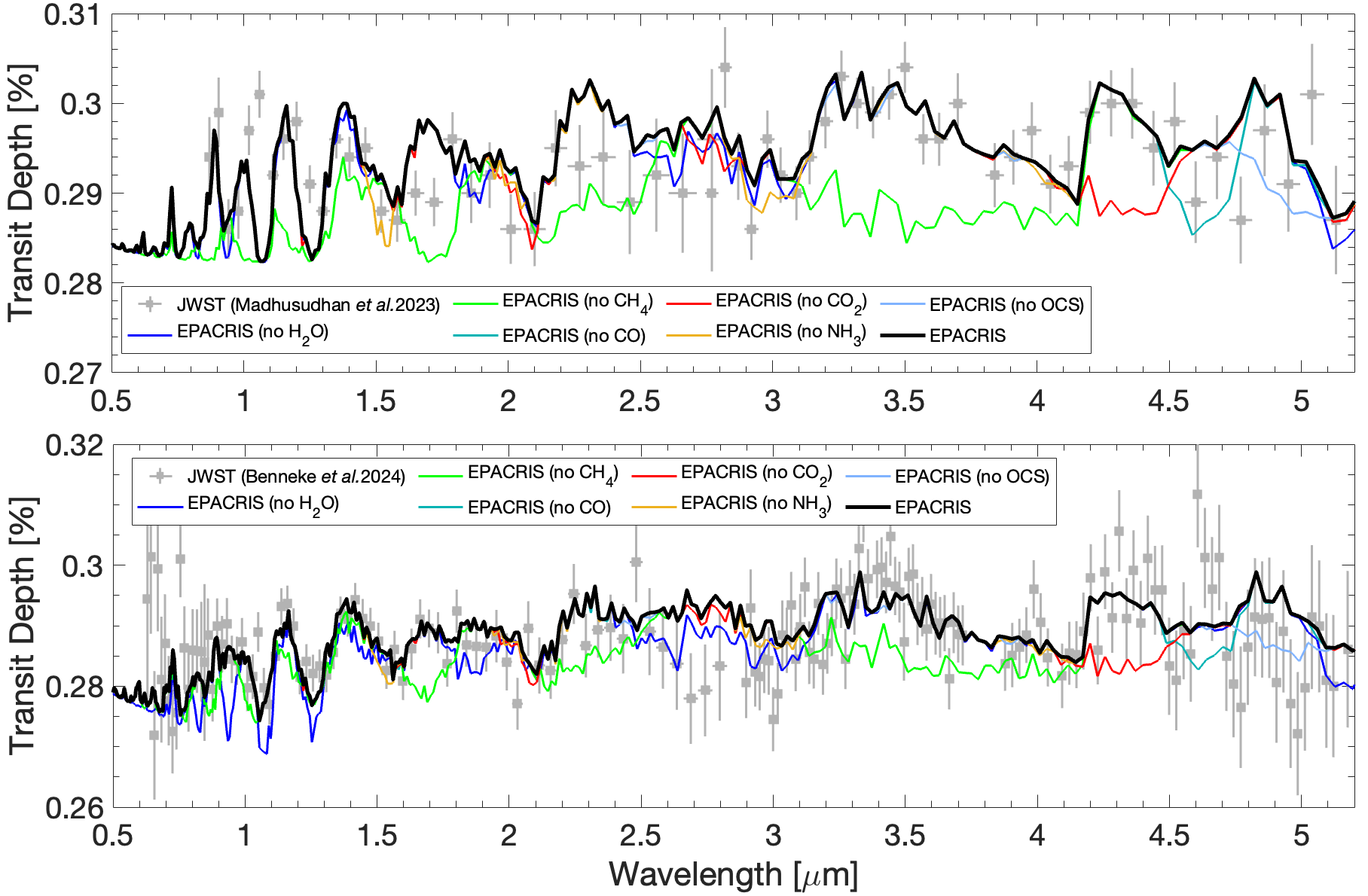}
    \caption{\footnotesize Comparisons between the theoretical transmission spectra generated by EPACRIS (solid lines) and the JWST observations for (Top) \ce{H2}:\ce{H2O}=50:50 and [C+N+S]=100$\times$solar metallicity scenario of K2-18~b assuming \textit{A}\textsubscript{b} = 0.3 corresponding to the solid lines in the left panel of Figure \ref{fig:vmr}, and (Bottom) \ce{H2}:\ce{H2O}=75:25 and [C+N+S]=100$\times$solar metallicity scenario of TOI-270~d assuming \textit{A}\textsubscript{b} = 0 corresponding to the solid lines in the right panel of Figure \ref{fig:vmr}. The grey symbols with error bars indicate JWST observations of corresponding exoplanets taken from \cite{Madhusudhan-2023} for K2-18~b (Top), and \cite{Benneke-2024_jwst} for TOI-270~d (Bottom). Each color represents a spectrum generated by excluding specific species: green for no \ce{CH4}, red for no \ce{CO2}, light blue for no \ce{OCS}, blue for no \ce{H2O}, teal for no \ce{CO}, light brown for no \ce{NH3}, and black for all species included.}
    \label{fig:transmission_spectra}
\end{figure*}

The bottom panel of Figure \ref{fig:transmission_spectra} compares EPACRIS-generated theoretical transmission spectra with JWST observations of TOI-270~d \citep{Benneke-2024_jwst}. The EPACRIS prediction generally aligns well with the JWST data, particularly in capturing the \ce{CO2} (red) and \ce{CH4} (green) features identified by \cite{Benneke-2024_jwst}. Unlike K2-18~b, water is present in the JWST observable part of TOI-270~d's atmosphere and contributes to the spectral features. Although a strong spectral feature around 4.7 $\mu$m cannot be solely explained by CO-attributed absorption, EPACRIS-generated transmission spectra suggest that the spectral modulation around 4.8$\mu$m is due to OCS. This, again, underscores the importance of detailed observations at this spectra range.

\subsection{Discussions}

In this study, we propose measuring the abundances of \ce{CO2} and \ce{CH4} (readily observable within 3--5 $\mu$m by JWST) as the primary metric to determine the bulk envelope \ce{H2O} to \ce{H2} ratio of temperate sub-Neptunes. Additionally, we suggest that other species such as CO, \ce{NH3}, OCS, and \ce{SO2} (see Section \ref{sec: species_mixing_ratio_as_a_constraint}) could provide further constraints on the envelope composition despite uncertainties in the deep-interior temperature. Notably, the inference of the envelope \ce{H2O} to \ce{H2} ratio does not involve direct measurements of \ce{H2O} in the observable part of the atmosphere, which can be depleted due to condensation.


Breaking the degeneracy in scenarios involving a hot interior is an intriguing question. In the recent study of warm-Neptune WASP-107~b (\textit{T}\textsubscript{eq}=750 K), the combination of \ce{CO2} and \ce{CH4}, which are jointly sensitive to \textit{T}\textsubscript{int}, was used to decipher the intrinsic temperature of a warm-Neptune exoplanet (\textit{T}\textsubscript{int}$\geq$ 400 K) \citep{sing_2024, welbanks_2024}. As mentioned in Section \ref{sec: co2toch4-ratio_as_a_constraint}, the model results are sensitive to \textit{T}\textsubscript{int} when the atmosphere is relatively \ce{H2}-dominated ($\leq$25\% water-rich) as shown in Figure \ref{fig:CO2_CH4_ratios} and \ref{fig:all_species}. Within the range of \textit{T}\textsubscript{int}$\leq$100 K, this sensitivity largely vanishes when the envelope is highly enriched in \ce{H2O}. It would be an interesting future study to determine if this insensitive to \textit{T}\textsubscript{int} in \ce{H2O}-enriched cases still applies when \textit{T}\textsubscript{int}$\geq$400 K (e.g., WASP-107~b; \cite{sing_2024,welbanks_2024}). Beyond the \ce{CO2}-\ce{CH4} abundances and ratio, our model suggests that the detection of \ce{SO2} in a relatively \ce{H2}-dominated atmosphere ($\leq$20\% internal \ce{H2O} envelope) of temperate sub-Neptunes could disfavor a high intrinsic temperature (\textit{T}\textsubscript{int}$\geq$100 K) scenario. 

We have identified various novel formation pathways for \ce{SO2} in sub-Neptune atmospheres. Unlike hot or warm-Jupiters or Neptunes, temperate sub-Neptunes around M-dwarf stars experience lower radiation on their planetary atmospheres, making the photochemical formation mechanism of \ce{SO2} proposed for the atmosphere of WASP-39~b ineffective \citep{Tsai_2023}. As shown in Section \ref{sec: species_mixing_ratio_as_a_constraint}, \ce{SO2} formation in temperate sub-Neptune atmospheres is complex due to many contributing factors, with a novel mechanism using \ce{CO2} as the oxidizer in weakly \ce{H2O}-rich atmospheres and direct thermochemical formation in highly \ce{H2O}-rich atmospheres. These mechanisms indicate that it is likely to detect \ce{SO2} on exoplanets with a wide range of envelope composition and temperatures, and the interpretation of each detection will likely require detailed photochemical modeling.

As shown in Figure B\ref{fig:TP-profiles}, the equilibrium temperature differences between K2-18~b and TOI-270~d distinctly categorize these planets based on whether water-condensation occurs (K2-18~b) or not (TOI-270~d). The equilibrium temperature of LP791-18~c falls between these two different regimes, presenting possible degenerate scenarios. Therefore, by performing chemical mapping of such temperate sub-Neptunes (e.g., LP791-18~c), we can cover a broad range of temperate sub-Neptunes with equilibrium temperatures (\textit{T}\textsubscript{eq}) between 250 and 400 K (corresponds to 221 confirmed sub-Neptunes according to \cite{NASA_exo_archive}). With the upcoming JWST observations of LP791-18~c, and many other future observations of temperate sub-Neptunes, the current chemical mapping will significantly enhance our understanding of a planet's interior O/H composition and potentially its planetary formation mechanisms as well.

\section{Conclusions} \label{sec:conclusions}
In this study, using self-consistent radiative transfer modeling \citep{Scheucher-2024} and a rate-based automatic chemical network generator combined with 1D photochemical kinetic-transport atmospheric modeling \citep{Yang_2024}, we have extensively investigated various atmospheric scenarios of temperate sub-Neptunes with equilibrium temperatures (\textit{T}\textsubscript{eq}) ranging from 250 to 400 K. We introduce a new framework that utilizes the atmospheric \ce{CO2}/\ce{CH4} ratio to gauge the bulk envelope \ce{H2O}/\ce{H2} ratio of temperate sub-Neptunes, providing new insights into their formation locations relative to the ice-line. Furthermore, our models suggest that any potential detection of OCS and \ce{SO2} serves as strong indicators of at least a 10\% water-rich envelope in temperate sub-Neptunes.

Benchmarking with recent JWST observations of two well-known temperate sub-Neptunes, K2-18b \citep{Madhusudhan-2023} and TOI-270d \citep{Benneke-2024_jwst}, our modeling results suggest the following: The most probable scenario for TOI-270d's envelope is \ce{H2}:\ce{H2O} $\sim$ 75:25 with [C+N+S] = 100$\times Z_{\odot}$, implying an original location inside the ice line during planetary formation. For K2-18b's envelope, the ratio is \ce{H2}:\ce{H2O} $\sim$ 50:50 with [C+N+S] $\leq$ 100$\times Z_{\odot}$, implying a location beyond the ice line during planetary formation. The synthesized transmission spectra based on these models showed good agreement with the JWST observations \citep{Madhusudhan-2023, Benneke-2024_jwst}. Furthermore, our synthesized transmission spectra suggest the potential detection of carbonyl sulfide (OCS) at approximately 4.8--4.9 $\mu$m in both JWST observations of K2-18~b and TOI-270~d. This indicates that these planets might contain at least a 10\% \ce{H2O}-rich envelope.

While TOI-270~d can be explained by conventional multipliers of solar metallicity, yielding approximately 230$\times Z_{\odot}$, K2-18~b cannot be solely explained by the conventional solar metallicity framework. This requires considering an additional variable: the deep-interior \ce{H2O}/\ce{H2} ratio, or planetary O/H ratio, which can potentially reveal the planets' original locations during their formation.

\section*{Acknowledgements}
This research work was carried out at the Jet Propulsion
Laboratory, California Institute of Technology, under a
contract with the National Aeronautics and Space Administration.
This research work was funded by the Caltech-JPL President's and Director's Research and Development Fund. © 2024. California Institute of Technology. Government sponsorship acknowledged.
%

\vspace{5mm}


\software{EPACRIS \citep{Hu_2012, Hu_2013, Hu_2014, Hu_2019, Yang_2024, Scheucher-2024}, RMG \citep{Gao_2016,rmg-v3,RMG-database, RMG-developers}}


\appendix
\setcounter{figure}{0} 

\section{elemental parameterization of planetary envelopes (Section 2.1)\label{sec:appendix_a}}
\begin{figure*}
    \renewcommand{\figurename}{Figure A}
    \centering    
    \includegraphics[width=\textwidth]{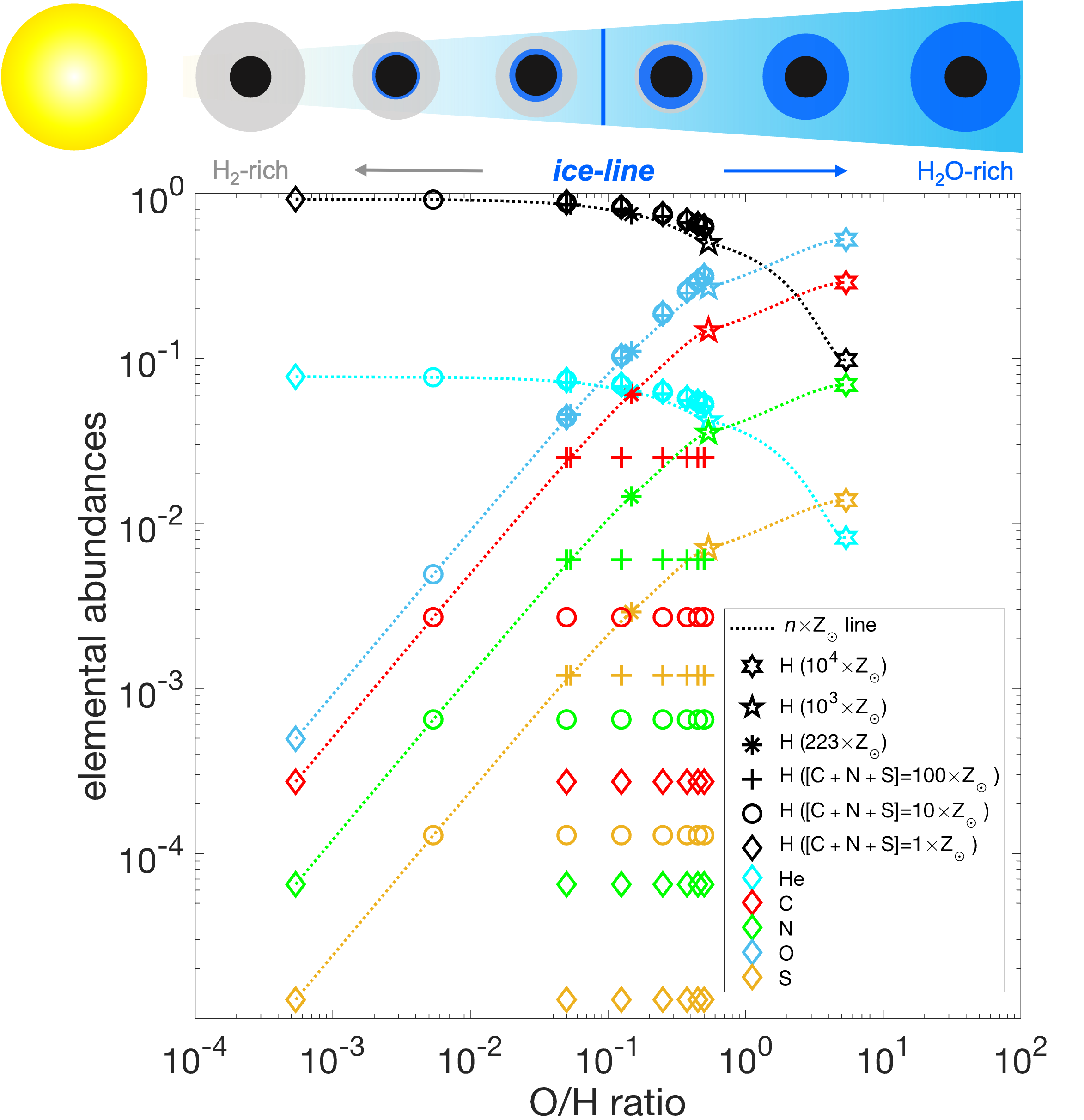}
     \caption{\footnotesize (Top) A schematic diagram of the protoplanetary disk with respect to the ice-line (note: the position of the ice-line is depicted arbitrarily to better illustrate the concept of the current study.); (Bottom) elemental abundances in various planetary envelope composition scenarios plotted against the Oxygen-to-Hydrogen (O/H) ratio, spanning from hydrogen-rich (i.e., lower O/H) to water-rich (i.e., higher O/H) envelopes. The O/H ratio along the X-axis represents the location within the protoplanetary disk relative to the ice-line, where planets with \ce{H2}-rich envelopes typically form inside the ice-line, and those with \ce{H2O}-rich envelopes form beyond the ice-line, as depicted in the top figure. Each open symbol corresponds to different elemental abundance scenarios ($Z_{\odot}$ represents solar metallicity): hexagrams for 10$^4\times Z_{\odot}$, asterisks for 10$^3\times Z_{\odot}$ solar metallicity, crosses for carbon, nitrogen, and sulfur abundances, [C+N+S], of 100$\times Z_{\odot}$, circles for [C+N+S] = 10$\times Z_{\odot}$, and diamonds for [C+N+S] = 1$\times Z_{\odot}$. Colors represent different elemental species: hydrogen (black), helium (cyan), carbon (red), nitrogen (green), oxygen (light blue), and sulfur (light brown). Dotted lines indicate standard multipliers of solar metallicity for each element.}  
    \label{fig:elemental_abundances}
\end{figure*}

This appendix section demonstrates a detailed description of the O/H variation mentioned in Section \ref{subsec:elements}. Figure A\ref{fig:elemental_abundances} and Table A\ref{tab:atom_abundances} present the various planetary envelope composition scenarios for sub-Neptunes as discussed in Section \ref{subsec:elements}. We began the O/H variation using the following equation:
\begin{gather*}
    [\ce{H}+\ce{He}+\ce{O}]_{n\times Z_{\odot}}=a\\
    \\
    \frac{[\ce{H2O}]}{[\ce{H2}]+[\ce{H2O}]}=x_{\text{acc}},\\
\end{gather*} 
where [\ce{H2O}] and [\ce{H2}] represent water and hydrogen abundance in the envelope, respectively. Therefore, $x_{\text{acc}}$ represents the accretion ratio for \ce{H2O}. If planetary formation takes place inside the ice line and closer to its parent star, $x_{\text{acc}}$ will approach 0 (i.e., \ce{H2}-rich). Conversely, if planetary formation occurs beyond the ice line and farther from its parent star, $x_{\text{acc}}$ will approach 1 (i.e., \ce{H2O}-rich).
Consequently, each elemental abundance [H], [O], and [He] will be as the following equation: 
\begin{equation*}
    \begin{split} 
     [\ce{H}]&=2([\ce{H2}]+[\ce{H2O}])\\
    [\ce{He}]&=\frac{[\ce{H}]}{11.91}=\frac{2([\ce{H2}]+[\ce{H2O}])}{11.91}\\
    [\ce{O}]&=[\ce{H2O}]=x_{\text{acc}}([\ce{H2}]+[\ce{H2O}]),\\
    \end{split}
\end{equation*}
where the H/He ratio was maintained at 11.91 to determine the helium abundance \citep{Lodders-2020}. Finally, the absolute elemental abundance for the $x_{\text{acc}}$ scenarios would be
\begin{equation}
    \begin{split}
    [\ce{H}]_{x_{\text{acc}}}&=a\times\frac{[\ce{H}]}{[\ce{H}]+[\ce{He}]+[\ce{O}]}\\
    &=\frac{2a}{2+x_{\text{acc}}+\frac{2}{11.91}}\\
    [\ce{He}]_{x_{\text{acc}}}&=a\times\frac{[\ce{He}]}{[\ce{H}]+[\ce{He}]+[\ce{O}]}\\
    &=\frac{2a}{ \left(2+x_{\text{acc}}+\frac{2}{11.91}\right)\times11.91}\\
        [\ce{O}]_{x_{\text{acc}}}&=a\times\frac{[\ce{O}]}{[\ce{H}]+[\ce{He}]+[\ce{O}]}\\
    &=\frac{ax_{\text{acc}}}{2+x_{\text{acc}}+\frac{2}{11.91}}\\
    \end{split}
    \label{eqn:elements}
\end{equation}

\begin{equation*}
    [\ce{H}+\ce{He}+\ce{O}]_{x_{\text{acc}}}=[\ce{H}+\ce{He}+\ce{O}]_{n\times Z_{\odot}}=a
\end{equation*}
Overall, the mixing ratio of [H+He+O] is unchanged, thus retaining the original ratios of carbon, nitrogen, and sulfur fixed.

As an example, for the \ce{H2}:\ce{H2O}=90:10 accretion scenario at 1$\times Z_{\odot}$ (the 2nd row and the 4th column in Table A\ref{tab:atom_abundances}), 
\begin{gather*}
    a=0.921775+0.077379+0.00495=1.004104\\
    \\
    x_{\text{acc}}=0.1\\
\end{gather*}
This results in a total H-to-O ratio of 20.00 (which can be approximated using 90$\times$2+10$\times$2 versus 10). Following the same procedure, we progressed through 75:25, 50:50, 25:75, and 10:90, culminating in a completely water-rich envelope at a 0:100 ratio. It should be noted that in the case of 100$\times Z_{\odot}$, if we assume all oxygen to be in the form of water, then the amount of water already exceeds the abundance of water in the \ce{H2}:\ce{H2O}=90:10 scenario. Consequently, the oxygen abundance at 100$\times Z_{\odot}$ (i.e.,\ce{H2}:\ce{H2O}=100:0) is slightly higher than the oxygen abundance at \ce{H2}:\ce{H2O}=90:10, as seen by comparing the third and fourth columns in the 20th row in Table A\ref{tab:atom_abundances}.

We applied these 21 scenarios (3 different solar metallicity cases $\times$ 7 different O/H ratio or \ce{H2}-to-\ce{H2O} accretion ratio) across three temperate sub-Neptunes: K2-18~b, LP791-18~c, and TOI-270~d, resulting in a total of 63 scenarios. We investigated 17 additional scenarios to understand model sensitivities. For K2-18~b, we tested the sensitivity of the \ce{H2}:\ce{H2O}=50:50 and [C+N+S]=100$\times Z_{\odot}$ scenario to the eddy diffusion coefficients (\textit{K}\textsubscript{zz}) with values of 10$^4$ and 10$^8$ cm\textsuperscript{2}/s, compared to the standard 10$^6$ cm\textsuperscript{2}/s mainly adopted in this study. We also explored 7 different O/H scenarios, maintaining [C+N+S] at 100 $\times Z_{\odot}$, using a \textit{T--P} profile computed for 100$\times Z_{\odot}$ and an intrinsic temperature (\textit{T}\textsubscript{int}) of 100 K to test the sensitivity of the model to a higher \textit{T}\textsubscript{int} of 100 K compared to the standard 60 K mainly adopted in this study. For TOI-270~d, we evaluated one scenario assuming 223$\times$ solar metallicity to benchmark against the metallicity constraint retrieved from \cite{Benneke-2024_jwst}. Additionally, we investigated 7 different O/H scenarios, maintaining [C+N+S] at 100$\times$ solar metallicity, assuming a Bond albedo of 0. Overall, we analyzed 74 (63+2+1+1+7) planetary atmospheric structures (\textit{T--P} profiles) and 80 (63+2+7+1+7)  planetary atmospheric photochemical models.

\begin{table*}[ht]
\renewcommand{\tablename}{Table A}
\caption{Elemental composition profiles utilized to investigate a range of planetary atmospheric scenarios in this study. We explored atmospheres spanning from hydrogen-rich (\ce{H2}) to water-rich (\ce{H2O}) compositions and varying from 1$\times$ to 100$\times$ solar metallicity. For each scenario, we systematically varied the \ce{H2} to \ce{H2O} ratio, beginning at 100:0 and progressing through intermediate stages of 90:10, 75:25, 50:50, 25:75, and 10:90, until reaching a completely water-rich environment at a 0:100 ratio. This range of scenarios, detailed in the main text and depicted in Figure A\ref{fig:elemental_abundances}, illustrates variations in planetary formation locations relative to the ice line. Additionally, some cases were modeled at higher metallicities, including 223$\times$, 1000$\times$, and 10000$\times$ solar metallicity.}\label{tab:atom_abundances}
\begin{tabular*}{\textwidth}{@{\extracolsep\fill} ccccccccc}
\toprule%
\multicolumn{2}{c}{\multirow{2}{*}{\shortstack{\textbf{\ce{H2}:\ce{H2O}}\\ \textbf{accretion ratio}}}} & \multicolumn{3}{c}{\textbf{\ce{H2}-rich}}  & \multicolumn{1}{c}{\textbf{$\Longleftrightarrow$}}  & \multicolumn{3}{c}{\textbf{\ce{H2O}-rich}}\\
\multicolumn{2}{c}{} & \textbf{100:0}\ & \textbf{90:10} & \textbf{75:25} & \textbf{50:50} & \textbf{25:75} & \textbf{10:90} & \textbf{0:100} \\

\midrule
\multirow{6}{*}{\rotatebox[origin=c]{0}{\textbf{1$\times$}}} & \textbf{H} & 0.921775 & 0.881567	& 0.826877 & 0.749393 & 0.685186 & 0.651685 & 0.631113\\
& \textbf{He} &0.077379	& 0.074004 & 0.069413 & 0.062909 & 0.057519 & 0.054706 & 0.052979 \\
& \textbf{C} & 0.000272	& 0.000272	& 0.000272 & 0.000272 & 0.000272 & 0.000272 & 0.000272\\
& \textbf{N} & 0.000065	& 0.000065	& 0.000065	& 0.000065	& 0.000065	& 0.000065 & 0.000065\\
& \textbf{O} & 0.000495	& 0.044078 & 0.103360 &	0.187348 &	0.256945 &	0.293258 &	0.315557 \\
& \textbf{S} & 0.000013	& 0.000013 & 0.000013& 0.000013 & 0.000013 & 0.000013 &	0.000013\\

\midrule
\multirow{6}{*}{\rotatebox[origin=c]{0}{\textbf{10$\times$}}} & \textbf{H} &  0.914815 & 0.878810 & 0.824291 & 0.747049 & 0.683043 & 0.649647 & 0.629140\\
& \textbf{He} & 0.076795 & 0.073773 & 0.069196 & 0.062712 & 0.057339 & 0.054535 & 0.052814\\
& \textbf{C} & 0.002700 & 0.002700 & 0.002700 & 0.002700 & 0.002700 & 0.002700 & 0.002700\\
& \textbf{N} & 0.000648 & 0.000648 & 0.000648 & 0.000648 & 0.000648 & 0.000648 & 0.000648\\
& \textbf{O} & 0.004913 & 0.043941 & 0.103036 & 0.186762 & 0.256141 & 0.292341 & 0.314570 \\
& \textbf{S} & 0.000129 & 0.000129 & 0.000129 & 0.000129 & 0.000129 & 0.000129 & 0.000129\\

\midrule
\multirow{6}{*}{\rotatebox[origin=c]{0}{\textbf{100$\times$}}} & \textbf{H} & 0.850591 & 0.853369 & 0.800428 & 0.725422 & 0.663269 & 0.630840 & 0.610926\\
& \textbf{He} & 0.071404 & 0.071637 &  0.067193 &  0.060896 & 	0.055679 & 0.052956 & 0.051285 \\
& \textbf{C} & 0.025103 & 0.025103 & 0.025103 & 0.025103 & 0.025103 & 0.025103 & 0.025103\\
& \textbf{N} & 0.006022 & 0.006022 & 0.006022 & 0.006022 & 0.006022 & 0.006022 & 0.006022 \\
& \textbf{O} & 0.045679 & 0.042668 & 0.100053 &	0.181356 & 0.248726 &	0.283878 & 0.305463\\
& \textbf{S} & 0.001201	 & 0.001201	& 0.001201 & 0.001201 &	0.001201 & 0.001201 & 0.001201\\
\midrule
\multirow{6}{*}{\rotatebox[origin=c]{0}{\textbf{223$\times$}}} & \textbf{H} & 0.748399 &  &  & &  &  & \\
& \textbf{He} & 0.062838 &  &  & &  &  &  \\
& \textbf{C} & 0.060745 &  &  & &  &  & \\
& \textbf{N} & 0.014572 &  &  & &  &  &  \\
& \textbf{O} & 0.110539 &  &  & &  &  & \\
& \textbf{S} & 0.002907	 &  &  & &  &  & \\
\midrule
\multirow{6}{*}{\rotatebox[origin=c]{0}{\textbf{1000$\times$}}} & \textbf{H} & 0.499741 &  &  & &  &  & \\
& \textbf{He} & 0.041960 &  &  & &  &  &  \\
& \textbf{C} & 0.147484 &  &  & &  &  & \\
& \textbf{N} & 0.035379 &  &  & &  &  &  \\
& \textbf{O} & 0.268377 &  &  & &  &  & \\
& \textbf{S} & 0.007059	 &  &  & &  &  & \\
\midrule
\multirow{6}{*}{\rotatebox[origin=c]{0}{\textbf{10000$\times$}}} & \textbf{H} & 0.097516 &  &  & &  &  & \\
& \textbf{He} & 0.008188 &  &  & &  &  &  \\
& \textbf{C} & 0.287791 &  &  & &  &  & \\
& \textbf{N} & 0.069036 &  &  & &  &  &  \\
& \textbf{O} & 0.523694 &  &  & &  &  & \\
& \textbf{S} & 0.013775	 &  &  & &  &  & \\
\botrule
\end{tabular*}
\end{table*}

\section{The temperature-pressure (T--P) profiles for various planetary envelope scenarios (Section 2.2)\label{sec:appendix_b}}
\begin{figure*}
    \renewcommand{\figurename}{Figure B}
    \centering    
    \includegraphics[width=0.8\textwidth]{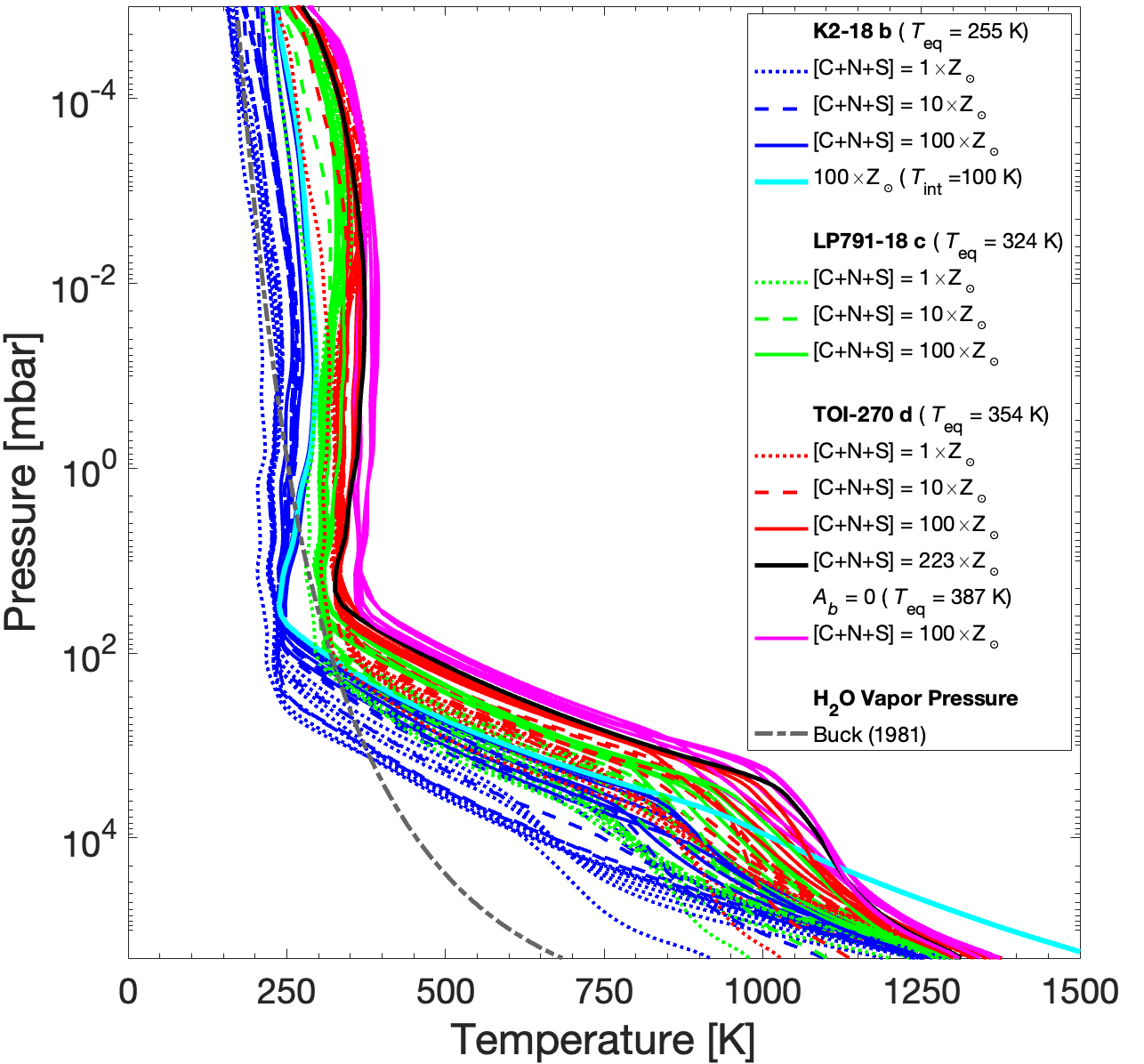}
    \caption{\footnotesize The figure shows 72 temperature-pressure \textit{T--P} profiles for different planetary envelope scenarios, calculated as described in Section \ref{sec:TP_profiles} (1000$\times$ and 10000$\times$ solar metallicity cases for K2-18~b are omitted for simplicity). $Z_{\odot}$ represents solar metallicity. Unless otherwise specified, the planetary equilibrium temperature (i.e., \textit{T}\textsubscript{eq}) assumes a \textit{A$_b$} of 0.3 and an intrinsic temperature (\textit{T}\textsubscript{int}) of 60 K by default. Each color corresponds to a specific planetary envelope scenario: blue for K2-18~b, cyan for K2-18~b assuming \textit{T}\textsubscript{int}=100 K, green for LP791-18~c, red for TOI-270~d, black for the 223$\times Z_{\odot}$ scenario of TOI-270~d, and magenta for TOI-270~d assuming \textit{A$_b$}=0. The solid lines (except cyan and black) represent seven different O/H scenarios (i.e., \ce{H2}:\ce{H2O} from 100:0 to 0:100) with the elemental composition of carbon, nitrogen, and sulfur ([C+N+S]) of 100$\times Z_{\odot}$, dashed lines for 10$\times Z_{\odot}$, and dotted lines for 1$\times Z_{\odot}$, as detailed in Section \ref{subsec:elements}. The grey solid-dashed line indicates the condensation curves for \ce{H2O} (solid and liquid states) from \cite{Buck-1981}. As illustrated in the figure, only for cases of K2-18~b (i.e., blue and cyan lines), water condenses to form clouds below the photosphere (i.e., the pressure range at $\sim$ 1 mbar where the JWST conducts its observations). In contrast, other planets, LP791-18~c (green lines) and TOI-270~d (red, black, and magenta lines), are too hot for water condensation to occur.}
    \label{fig:TP-profiles}
\end{figure*}

This appendix section provides detailed information on the radiative transfer modeling using EPACRIS-Climate \citep{Scheucher-2024} as described in Section \ref{sec:TP_profiles}. Figure B\ref{fig:TP-profiles} represents the resulting temperature-pressure \textit{T--P} profiles calculated for different planetary envelope scenarios.

EPACRIS-Climate \citep{Scheucher-2024} uses the two-stream method \citep{heng2018radiative} to compute radiative fluxes and incorporates both dry and moist adiabatic adjustments, following the methodology of \cite{graham2021multispecies}. In brief, the climate module of EPACRIS automatically applies moist adiabats when condensation occurs, and dry adiabats when condensation does not occur. Water is considered a condensable component, and its atmospheric concentration is self-consistently adjusted in line with the moist adiabats. Some might argue that in an \ce{H2} background, moist convection is inhibited once water surpasses a critical moisture threshold, resulting in a steep lapse rate \citep{Leconte-2017, Innes-2023, Leconte-2024}. Also, in the interior environment of the interior where \textit{T}$\geq\sim$650 K \textit{P}$\geq\sim$200 bar, water becomes supercritical and is expected to be miscible with other gases \citep{Pierrehumbert-2023, Benneke-2024_jwst}. Although the current study does not model the deep interior beyond 200 bars, the climate module of EPACRIS currently does not account for supercritical fluid adiabats. Therefore, addressing these features could be a potential future development in the climate module of EPACRIS. Presently, EPACRIS considers the impact of a hotter interior on chemical composition, incorporating a higher internal heat flux. 

We assumed an intrinsic temperature (\textit{T}\textsubscript{int}) of 60 K \citep{Hu_2021}. For some cases of K2-18~b scenarios, we tested larger \textit{T}\textsubscript{int} of 100 K to test the sensitivity of the model, since \textit{T}\textsubscript{int} can impact deep atmospheric quenching and its result in the upper atmospheric abundances of chemical species \citep{Fortney_2020, Tsai_2021}. 

By default, we set the planetary Bond albedo (\textit{A}$_b$) to an Earth-like value of 0.3 for all planets, including K2-18~b, LP791-18~c, and TOI-270~d. For TOI-270d, we additionally explored seven different O/H ratio scenarios while maintaining [C+N+S]=100$\times Z_{\odot}$, but with \textit{A}$_b$=0. In all cases, we assumed an average solar zenith angle of 48.19$^{\circ}$ \citep{Cronin-2014}. The stellar irradiation parameters ($S_p$ in W$\cdot$ m$^{-2}$) were adopted to be 1368 W$\cdot$ m$^{-2}$ for K2-18~b (\cite{Benneke_2019}), 3607 W$\cdot$ m$^{-2}$ for LP 791-18~c (\cite{peterson-2023}), and 4333 W$\cdot$ m$^{-2}$ for TOI-270~d (\cite{gunther-2019}).

\section{RMG-generated species that are not included in the original EPACRIS species library (Section 2.3)\label{sec:appendix_c}}
This appendix section provides additional information on the chemical reaction network generation using RMG \citep{Gao_2016} as described in Section \ref{sec:rmg}, and details the 36 newly included molecular species in the chemical network tailored for \ce{H2O}-dominated atmospheres.

In this work, we sampled temperatures ranging from 300 to 1100 K and pressures from 10\textsuperscript{-3} to 10\textsuperscript{2} bar to generate chemical networks tailored for a \ce{H2O}-rich scenario of temperate sub-Neptunes, as established by previous radiative transfer modeling of K2-18~b \citep{Scheucher_2020}. 

For initial molecular mixing ratio inputs required for using RMG to generate a reaction network for the targeted system, we used the following: $10^{0}$ for \ce{H2O} to account for a \ce{H2O}-rich scenario. Additional constraints were based on the recent JWST observations of K2-18~b by Madhusudhan (2023): $10^{-1.89}$ for \ce{CH4} and $10^{-2.05}$ for \ce{CO2}, both determined from two-offsets retrieved molecular abundances with detection significances of 5$\sigma$ and 3.2$\sigma$, respectively; $10^{-4.46}$ for \ce{S} and \ce{N}, derived from no-offset retrieved molecular abundances with a 2.4$\sigma$ detection significance for \ce{(CH3)2S} and based on the upper limit for nitrogen. These values were later normalized to sum to 1.

The choice of reaction libraries can be found in the supplementary RMG input file. The details of these libraries can be found in the RMG database \cite{RMG-developers}. The completed network comprised 92 species and 1950 reactions, detailed in the supplementary CHEMKIN file. Of these, 92 species and 1666 reactions were incorporated into the 1-D photochemical kinetic-transport modeling after the treatment described in \cite{Yang_2024}. 

Among 92 species, 36 species are newly generated by RMG, and not available in the EPACRIS library. When compared to other well-known photochemical networks used for exoplanetary species, within these 36 newly generated species by RMG, 27 species were not included in KINETICS \citep{Moses_2011}, and 30 species were not included in VULCAN \citep{Tsai_2017, Tsai_2021} (see Table C\ref{tbl: newspecies} for detail). Among these not-included species, hydroxymethylene (HCOH) is shown to be a major intermediate species for the conversion of carbon monoxide into methane and water \citep{Yang-2023}. Additionally, many hydroxyl group species such as \ce{CH2CHOH}, \ce{CH2CH2OH}, \ce{HCCOH}, \ce{CHCHOH}, \ce{CH3CH2OH}, and \ce{CH3CHOH} participate in the dehydration reactions that link simple molecules into long and complex chains. For this reason, although these molecules may be challenging to detect with current observational techniques, including them in photochemical modeling is essential, as this can potentially lead to the formation of novel species. 

To make sure the chemical reaction network generated for the \ce{H2O}-rich conditions are also applicable to the $x$\textsubscript{acc}=0 scenario (i.e., \ce{H2}:\ce{H2O}=100:0), we conducted photochemical modeling of the \ce{H2}-dominated K2-18~b atmosphere using the same atomic abundance and the same \textit{T--P} profile adopted in MODEL 3 of \cite{wogan2024jwst}. As discussed in Section \ref{sec: co2toch4-ratio_as_a_constraint}, the predicted mixing ratios of major species such as \ce{CO2}, \ce{CH4}, \ce{H2O}, \ce{N2}, \ce{CO}, \ce{NH3}, and \ce{SO2} were almost identical, confirming the robustness of the model results in both the \ce{H2O}- and \ce{H2}-dominated regimes.

\begin{table}
\renewcommand{\tablename}{Table C}
\caption{36 newly included molecular species in the chemical network tailored for \ce{H2O}-dominated atmospheres by RMG}
\centering
\begin{tabular}{|ll|ll|}
\hline
\textbf{Species} & \textbf{SMILES}\textsuperscript{\textit{a}} & \textbf{Species} & \textbf{SMILES}\textsuperscript{\textit{a}}\\ \hline
   \ce{CH2OH}\textsuperscript{\textit{b,c}} & [CH2]O&\ce{HCOH}& [C-]=[OH+] \\ \hline
   \ce{CH2CHO} &[CH2]C=O&\ce{H2CC}&[C]=C\\ \hline
   \ce{CH2CHOH} &C=CO&\ce{CH2CH2OH}&[CH2]CO\\ \hline
   \ce{CHCHO} &[CH]=C[O]&\ce{HCCOH}&C\#CO\\ \hline
   \ce{CHCHOH}&[CH]=CO&\ce{CH3CH2OH}&CCO\\ \hline
   \ce{CH3CHOH} &C[CH]O&\ce{OCHCO}&O=[C]C=O\\ \hline
   \ce{HOCH2O} &[O]CO&\ce{OCHO}&[O]C=O\\ \hline
   \ce{CH3C(O)O} &CC([O])=O&\ce{HSO2}&O=[SH]=O\\ \hline
   \ce{HOS} &O[S]&\ce{Sa}\textsuperscript{\textit{c}}&[S]\textsuperscript{\textit{d}}\\ \hline
   \ce{HSS}\textsuperscript{\textit{b,c}}&S=[SH]&\ce{HSSH}\textsuperscript{\textit{c}}&SS\\ \hline
   \ce{CH2SH}&[CH2]S&\ce{HCCS}&[S]C\#C\\ \hline   
   \ce{H2SS} &S=[SH2]&\ce{NCN}&[N]=C=[N]\\ \hline
   \ce{CH2N2} &[CH2]N=[N]&\ce{H2CN}\textsuperscript{\textit{b,c}}&C=[N]\\ \hline
   \ce{cNCN} &C1=NN=1&\ce{CH2NCN}&[CH2]N=C=[N]\\ \hline
   \ce{CH2NH}\textsuperscript{\textit{b,c}} &C=N&\ce{NCOH}&OC\#N\\ \hline
   \ce{NNH}\textsuperscript{\textit{b,c}}&[N]=N&\ce{CH3NH}\textsuperscript{\textit{c}}&C[NH]\\ \hline    
   \ce{CH2NNH} &[CH2]N=N&\ce{CH2NNH2}&NN=C\\ \hline
   \ce{CH2NH2}\textsuperscript{\textit{b,c}}&[CH2]N&\ce{CHNH}&[CH]=N\\ \hline
   
\end{tabular}
\label{tbl: newspecies}
\tablecomments{\footnotesize\textsuperscript{\textit{a}} Simplified Molecular-Input Line-Entry System\\ \textsuperscript{\textit{b}} species included in VULCAN \citep{Tsai_2017, Tsai_2021} \\ \textsuperscript{\textit{c}} species included in KINETICS \citep{Moses_2011} \\ \textsuperscript{\textit{d}} Despite appearing as doublet or triplet radicals in SMILES representation, these species are singlets in the `adjacency lists' representation, indicating all electrons are paired.}
\end{table}

\section{The eddy diffusion coefficients and stellar fluxes (Section 2.4)\label{sec:appendix_d}}
This appendix section provides additional information on the eddy diffusion coefficient profile and stellar fluxes used for 1-D photochemical kinetic-transport atmospheric modeling described in Section \ref{sec:epacris}

The eddy diffusion coefficient profile was estimated to be uniform 10$^6$ [cm$^2\cdot$s$^{-1}$] at various altitudes based on Figure 1 in \cite{Zhang_2018}. This value is comparable to that of Earth, which ranges from 4$\times10^3$--10$^6$ [cm$^2\cdot$s$^{-1}$], and Earth has planetary equilibrium temperature (\textit{T}\textsubscript{eq}) of 255 K, similar to K2-18~b. However, some might argue that \textit{K}\textsubscript{zz} could be much smaller, especially if a short-wavelength absorber causes a temperature inversion, as it is in Earth's stratosphere. To test the sensitivity of the steady-state vertical mixing ratios to changes in the eddy diffusion coefficients, we conducted additional modelings with coefficients set at 10$^4$ and 10$^8$ [cm$^2\cdot$s$^{-1}$] in selected cases. These tests showed that the results were not highly sensitive to variations in the eddy diffusion coefficient, which will be discussed again later.

For simulations involving K2-18 (classified as an M2.8-type star by \cite{Montet_2015}) and LP 791-18 (classified as an M6-type star by \cite{Crossfield_2019}), we scaled for the bolometric luminosity using the M5-type stellar spectrum of GJ876 from the MUSCLES survey III \citep{Loyd_2016}. For simulations involving TOI-270 (classified as an M3V-type star by \cite{gunther-2019}), the M3V-type stellar spectrum of GJ644 from the same survey was used for the bolometric luminosity scaling.

\section{Details of \ce{SO2} formation regimes (Section 3.1.2)\label{sec:appendix_e}}

In the regimes with less than approximately $\sim$20\% \ce{H2O} abundance inside the envelope, \ce{SO2} forms though the following reactions at \textit{P}$\sim$10 mbar:
\begin{equation}
    \begin{split}
        \ce{H2S}&\rightarrow\ce{H2}+\ce{S}\\
        \ce{S}+\ce{CO}&\rightarrow\ce{OCS}\\
        \ce{OCS}+\ce{S}&\rightarrow\ce{CO}+\ce{S2}\\
        \ce{S2}&\xrightarrow{\text{h$\nu$}}\ce{S}+\ce{S}\\
        \ce{S}+\ce{CO2}&\rightarrow\ce{SO}+\ce{CO}\\
        \ce{SO}+\ce{SO}&\rightarrow\ce{SO2}+\ce{S}\\
        \end{split}
    \label{eqn:SO2_CO2_driven}
\end{equation}
In this regime, \ce{CO2} acts as an oxidizer for sulfur due to its thermochemical instability as a carbon-bearing form (with \ce{CH4} being more favorable). Thus, the amount of \ce{SO2} formed is determined by the amount of \ce{CO2} (see mixing ratio patterns at deep interior \ce{H2O}/\ce{H2}$\leq$0.4 in both Figures \ref{fig:all_species}). Additional \ce{SO2} formation occurs in the upper atmosphere via photochemistry:
\begin{equation}
    \begin{split}        
        \ce{H2O}&\xrightarrow{\text{h$\nu$}}\ce{H}+\ce{OH}\\
        \ce{S}+\ce{OH}&\rightarrow\ce{SO}+\ce{H}\\
        \ce{SO}+\ce{OH}&\rightarrow\ce{SO2}+\ce{H}\\
    \end{split}
    \label{eqn:SO2_photochemistry}
\end{equation}
This is the proposed photochemical formation pathway for \ce{SO2} in the hot Jupiter-type exoplanet WASP-39~b \citep{Tsai_2023}.

The transition region between the $\leq$20\% \ce{H2O} region and the $\geq$20\% region turned out to be very sensitive to the \textit{T--P} structure of the planet. For example, when comparing the 100$\times Z_{\odot}$ scenario of K2-18~b assuming \textit{T}\textsubscript{int} = 60 K and 100 K, there is more than a 6 orders of magnitude difference between the predicted \ce{SO2} mixing ratios. This difference primarily arises from water condensation, which impacts the stability of \ce{H2S}, the main precursor for sulfur-bearing species such as \ce{SO2} and OCS. In the scenario where \textit{T}\textsubscript{int} = 100 K, the deep interior at 100 bar is approximately 250 K hotter compared to the interior assuming \textit{T}\textsubscript{int} = 60 K. At pressures lower than approximately 3 bar, the entire \textit{T--P} profile assuming \textit{T}\textsubscript{int} = 100 K is about 5 K higher compared to the \textit{T--P} profile assuming \textit{T}\textsubscript{int} = 60 K. This small difference of about 5 K results in about 15\% more \ce{H2O} abundance at \textit{P}$\leq\sim$3 bar for the \ce{H2}-dominated atmosphere of K2-18~b assuming \textit{T}\textsubscript{int} = 100 K, compared to when assuming \textit{T}\textsubscript{int} = 60 K. This 15\% excess \ce{H2O} favors the retention of \ce{H2S} through the reaction:
\begin{equation}
    \begin{split}        
        \ce{H2O}+\ce{SH}&\rightarrow\ce{H2S}+\ce{OH}\\
    \end{split}
    \label{eqn:H2O_reducing_H2S}
\end{equation}
As a result, in the case of \ce{H2}-dominated K2-18~b assuming \textit{T}\textsubscript{int} = 100 K, \ce{H2S} can remain stable up to \textit{P}$\sim$0.2 mbar, where \ce{H2S} photodissociates to form various sulfur-bearing species including \ce{SO2}. In contrast, with \textit{T}\textsubscript{int} = 60 K, \ce{H2S} starts to thermally dissociate into sulfur-bearing species already starting at \textit{P}$\sim$30 mbar. This results in significant differences between the predicted \ce{SO2} mixing ratio, as shown in the \ce{SO2} panels of Figure \ref{fig:all_species}.
This is also consistent with the OCS case, as evidenced by the rapid decrease of the OCS mixing ratio within this transition region (see blue crosses and blue hexagrams near 100$\times Z_{\odot}$ grey dotted line in Figure \ref{fig:all_species}). This \textit{T--P} structure-sensitive behavior of \ce{SO2} driven by \ce{H2O}-condensation implies that the detection of \ce{SO2} in temperate sub-Neptunes with an internal \ce{H2O} envelope of less than 20\% may be inconsistent with a high intrinsic temperature (\textit{T}\textsubscript{int}). 

In the \ce{H2O}-rich regime (more than approximately 20\% \ce{H2O} abundance inside the envelope), \ce{SO2} becomes thermochemically favorable in an oxidizing deep interior. For this reason, \ce{SO2} already exists in significant amounts--up to 100 ppm--in the deep interior, and is transported to the upper layers. In the upper atmosphere, additional \ce{SO2} is formed through Scheme \ref{eqn:SO2_photochemistry}. Although the vertical mixing ratio of \ce{SO2} varies depending on the eddy diffusion coefficients used in the model and cannot solely represent the predicted \ce{SO2} amount in the observable regime (i.e., \textit{P}$\sim$1 mbar) shown in Figure \ref{fig:all_species}, our model indicates that the thermochemical formation of \ce{SO2} in the deep-interior contributes significantly to the \ce{SO2} abundance if the interior is water-rich. In these conditions, \ce{SO2} becomes abundant with higher water content and deep-interior temperatures. As shown in Figure \ref{fig:all_species}, \ce{SO2} formation correlates positively with deep-interior temperature and the planetary envelope O/H ratio (i.e., \ce{H2O}/\ce{H2}). For instance, the deep-interior temperature tends to increase with the planetary equilibrium temperature. Thus, K2-18b has the lowest equilibrium temperature at 255 K, followed by LP 791-18c (324 K), TOI-270d (354 K), and TOI-270d assuming \textit{A}\textsubscript{b} = 0 with the highest equilibrium temperature of 387 K. Higher [C+N+S] abundance also promotes \ce{SO2} formation as well.

\section{Details of sulfur chemistry in the atmosphere of K2-18~b (Section 3.2.1)\label{sec:appendix_f}}

The summarized Scheme \ref{eqn:OCS_deep_interior_formation} for the deep-interior OCS formation is consistent with the similar vertical mixing ratio profile shape between OCS and CO at \textit{P}$\geq10^4$ mbar as shown in the left panel of Figure \ref{fig:vmr}. OCS is then transported upward via vertical mixing to the upper atmosphere (\textit{P}$\leq$10 mbar). In the upper atmosphere, additional OCS forms through a disequilibrium process, primarily driven by UV-photochemistry, as follows:
\begin{equation}
    \begin{split}
        \ce{H2S}&\xrightarrow{\text{h$\nu$}}\ce{SH}+\ce{H}\\
        \ce{SH}&\xrightarrow{\text{h$\nu$}}\ce{S}+\ce{H}\\
        \ce{S}+\ce{CO}&\xrightarrow{\text{M}}\ce{OCS}\\
    \end{split}
    \label{eqn:OCS_photochemical_formation}
\end{equation}
with M representing any third-body molecule.
This process aligns with the observed decrease in CO mixing ratios at \textit{P}$=10^{-4}-10^{-2}$ mbar, above which CO is replenished by \ce{CO2} UV-photodissociation. It should be noted that the OCS formation pathways here do not directly involve \ce{H2O} or its photolysis. Consequently, the OCS abundance is not sensitive to the condensation of water in the upper atmosphere.

Although sulfur dioxide (\ce{SO2}) is also predicted to form at higher altitudes (i.e., \textit{P}$=10^{-4}-10^{-3}$ mbar) through a similar photochemical scheme described in \cite{Tsai_2023}, its predicted abundance is insignificant to appear as notable spectral features, compared to the \ce{SO2} spectral feature observed in WASP-39~b \citep{Alderson-2023, Rustamkulov-2023, Ahrer-2023, Feinstein-2023, Powell_2024}.
\begin{figure*}
    \renewcommand{\figurename}{Figure F}
    \centering    
    \includegraphics[width=1\textwidth]{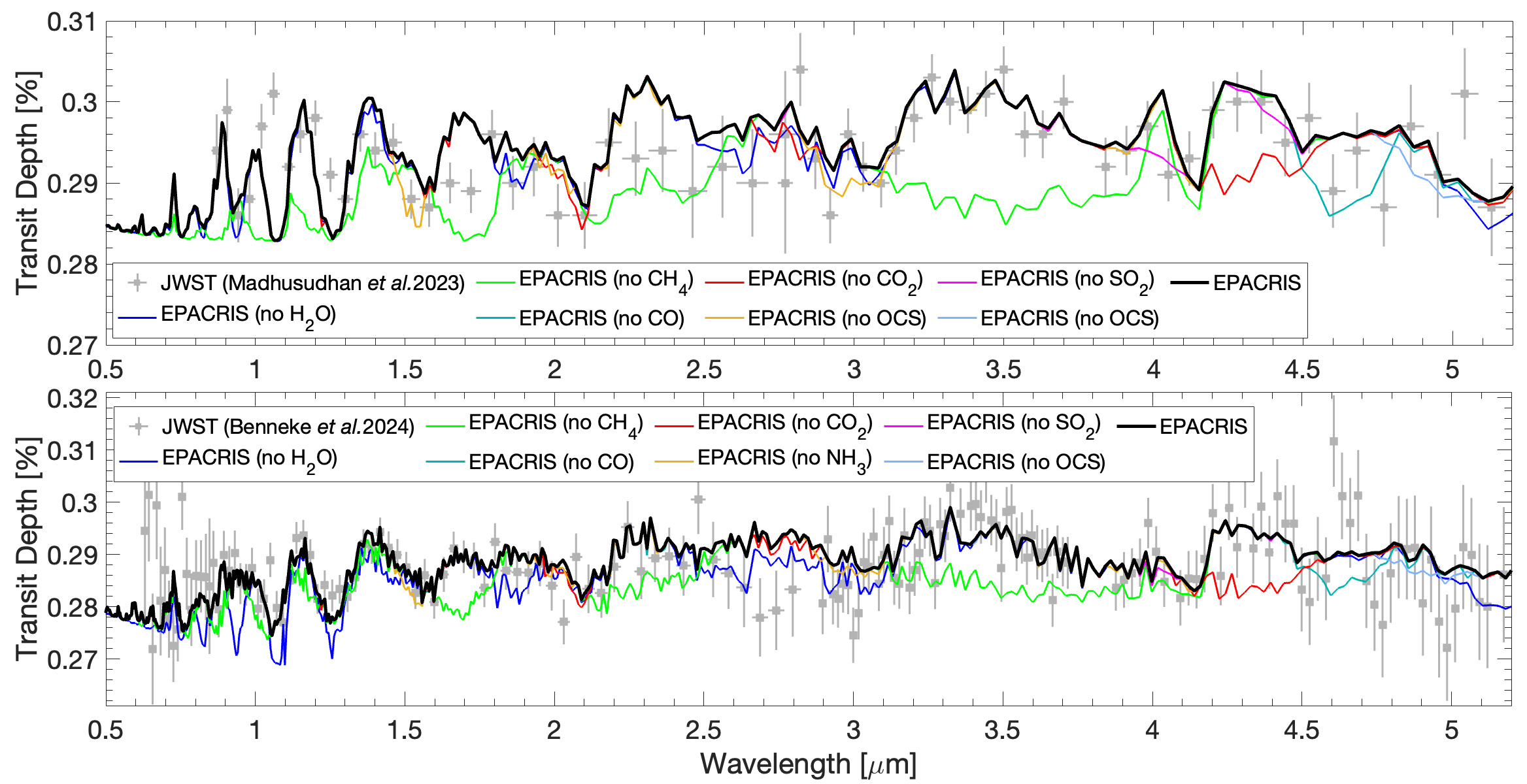}
    \caption{\footnotesize Comparisons between the theoretical transmission spectra generated by EPACRIS (solid lines) and the JWST observations for (Top) \ce{H2}:\ce{H2O}=50:50 and [C+N+S]=100$\times$solar metallicity scenario of K2-18~b assuming \textit{A}\textsubscript{b} = 0.3 corresponding to the dotted lines in the left panel of Figure \ref{fig:vmr}, and (Bottom) \ce{H2}:\ce{H2O}=75:25 and [C+N+S]=100$\times$solar metallicity scenario of TOI-270~d assuming \textit{A}\textsubscript{b} = 0 corresponding to the dotted lines in the right panel of Figure \ref{fig:vmr}. In this EPACRIS model prediction, the rate coefficient of Scheme \ref{eqn:OCS_recombination} is adopted from \cite{Tsai_2021}, and the major difference from Figure \ref{fig:transmission_spectra} is the amplitude of \ce{SO2} (significantly increased in the K2-18~b case) and OCS features (significantly decreased in the K2-18~b case). The grey symbols with error bars indicate JWST observations of corresponding exoplanets taken from \cite{Madhusudhan-2023} for K2-18~b (Top), and \cite{Benneke-2024_jwst} for TOI-270~d (Bottom). Each color represents a spectrum generated by excluding specific species: green for no \ce{CH4}, red for no \ce{CO2}, light blue for no \ce{OCS}, blue for no \ce{H2O}, teal for no \ce{CO}, light brown for no \ce{NH3}, and black for all species included.}
    \label{fig:transmission_spectra_OCS_recombination}
\end{figure*}

Figure F\ref{fig:transmission_spectra_OCS_recombination} shows the synthesized transmission spectra for K2-18~b (upper panel) and TOI-270~d (bottom panel) based on the dotted lines from Figure \ref{fig:vmr}. Most of the spectral features remain consistent, while the spectral features at $\sim$4 $\mu m$ and 4.8--5 $\mu m$ change, attributed to \ce{SO2} and \ce{OCS}. As mentioned in Section \ref{subsec:K2-18b_chemistry} of the main text, the community lacks extensive details on sulfur chemistry, and this strongly suggests the need for future studies to accurately estimate the rate coefficients for sulfur chemistry through both ab initio calculations and experimental reaction kinetic measurements, including Scheme \ref{eqn:OCS_recombination} and others, such as S(\textsuperscript{1}D) photochemistry.

\bibliography{reference}{}

\begin{thebibliography}{}
\expandafter\ifx\csname natexlab\endcsname\relax\def\natexlab#1{#1}\fi
\providecommand{\url}[1]{\href{#1}{#1}}
\providecommand{\dodoi}[1]{doi:~\href{http://doi.org/#1}{\nolinkurl{#1}}}
\providecommand{\doeprint}[1]{\href{http://ascl.net/#1}{\nolinkurl{http://ascl.net/#1}}}
\providecommand{\doarXiv}[1]{\href{https://arxiv.org/abs/#1}{\nolinkurl{https://arxiv.org/abs/#1}}}

\bibitem[{{Ahrer} {et~al.}(2023){Ahrer}, {Stevenson}, {Mansfield}, {Moran}, {Brande}, {Morello}, {Murray}, {Nikolov}, {Petit dit de la Roche}, {Schlawin}, {Wheatley}, {Zieba}, {Batalha}, {Damiano}, {Goyal}, {Lendl}, {Lothringer}, {Mukherjee}, {Ohno}, {Batalha}, {Battley}, {Bean}, {Beatty}, {Benneke}, {Berta-Thompson}, {Carter}, {Cubillos}, {Daylan}, {Espinoza}, {Gao}, {Gibson}, {Gill}, {Harrington}, {Hu}, {Kreidberg}, {Lewis}, {Line}, {L{\'o}pez-Morales}, {Parmentier}, {Powell}, {Sing}, {Tsai}, {Wakeford}, {Welbanks}, {Alam}, {Alderson}, {Allen}, {Anderson}, {Barstow}, {Bayliss}, {Bell}, {Blecic}, {Bryant}, {Burleigh}, {Carone}, {Casewell}, {Changeat}, {Chubb}, {Crossfield}, {Crouzet}, {Decin}, {D{\'e}sert}, {Feinstein}, {Flagg}, {Fortney}, {Gizis}, {Heng}, {Iro}, {Kempton}, {Kendrew}, {Kirk}, {Knutson}, {Komacek}, {Lagage}, {Leconte}, {Lustig-Yaeger}, {MacDonald}, {Mancini}, {May}, {Mayne}, {Miguel}, {Mikal-Evans}, {Molaverdikhani}, {Palle}, {Piaulet}, {Rackham}, {Redfield}, {Rogers}, {Roy}, {Rustamkulov},
  {Shkolnik}, {Sotzen}, {Taylor}, {Tremblin}, {Tucker}, {Turner}, {de Val-Borro}, {Venot}, \& {Zhang}}]{Ahrer-2023}
{Ahrer}, E.-M., {Stevenson}, K.~B., {Mansfield}, M., {et~al.} 2023, \nat, 614, 653, \dodoi{10.1038/s41586-022-05590-4}

\bibitem[{Alderson {et~al.}(2023)Alderson, Wakeford, Alam, Batalha, Lothringer, Redai, \& Barat}]{Alderson-2023}
Alderson, L., Wakeford, H.~R., Alam, M.~K., {et~al.} 2023, Nature

\bibitem[{Benneke {et~al.}(2019)Benneke, Wong, Piaulet, Knutson, Lothringer, Morley, Crossfield, Gao, Greene, Dressing, Dragomir, Howard, McCullough, Kempton, Fortney, \& Fraine}]{Benneke_2019}
Benneke, B., Wong, I., Piaulet, C., {et~al.} 2019, The Astrophysical Journal Letters, 887, L14, \dodoi{10.3847/2041-8213/ab59dc}

\bibitem[{Benneke {et~al.}(2024)Benneke, Roy, Coulombe, Radica, Piaulet, Ahrer, Pierrehumbert, Krissansen-Totton, Schlichting, Hu, Yang, Christie, Thorngren, Young, Pelletier, Knutson, Miguel, Evans-Soma, Dorn, Gagnebin, Fortney, Komacek, MacDonald, Raul, Cloutier, Acuna, Lafrenière, Cadieux, Doyon, Welbanks, \& Allart}]{Benneke-2024_jwst}
Benneke, B., Roy, P.-A., Coulombe, L.-P., {et~al.} 2024, JWST Reveals CH$_4$, CO$_2$, and H$_2$O in a Metal-rich Miscible Atmosphere on a Two-Earth-Radius Exoplanet.
\newblock \doarXiv{2403.03325}

\bibitem[{Brande {et~al.}(2024)Brande, Crossfield, Kreidberg, Morley, Barman, Benneke, Christiansen, Dragomir, Fortney, Greene, Hardegree-Ullman, Howard, Knutson, Lothringer, \& Mikal-Evans}]{Brande_2024}
Brande, J., Crossfield, I. J.~M., Kreidberg, L., {et~al.} 2024, The Astrophysical Journal Letters, 961, L23, \dodoi{10.3847/2041-8213/ad1b5c}

\bibitem[{Buck(1981)}]{Buck-1981}
Buck, A.~L. 1981, Journal of Applied Meteorology and Climatology, 20, 1527 , \dodoi{10.1175/1520-0450(1981)020<1527:NEFCVP>2.0.CO;2}

\bibitem[{Burn {et~al.}(2024)Burn, Mordasini, Mishra, Haldemann, Venturini, Emsenhuber, \& Henning}]{Burn-2024}
Burn, R., Mordasini, C., Mishra, L., {et~al.} 2024, Nature Astronomy, 8, 463

\bibitem[{Cronin(2014)}]{Cronin-2014}
Cronin, T.~W. 2014, Journal of the Atmospheric Sciences, 71, 2994

\bibitem[{Crossfield {et~al.}(2019)Crossfield, Waalkes, Newton, Narita, Muirhead, Ment, Matthews, Kraus, Kostov, Kosiarek, Kane, Isaacson, Halverson, Gonzales, Everett, Dragomir, Collins, Chontos, Berardo, Winters, Winn, Scott, Rojas-Ayala, Rizzuto, Petigura, Peterson, Mocnik, Mikal-Evans, Mehrle, Matson, Kuzuhara, Irwin, Huber, Huang, Howell, Howard, Hirano, Fulton, Dupuy, Dressing, Dalba, Charbonneau, Burt, Berta-Thompson, Benneke, Watanabe, Twicken, Tamura, Schlieder, Seager, Rose, Ricker, Quintana, Lépine, Latham, Kotani, Jenkins, Hori, Colon, \& Caldwell}]{Crossfield_2019}
Crossfield, I. J.~M., Waalkes, W., Newton, E.~R., {et~al.} 2019, The Astrophysical Journal Letters, 883, L16, \dodoi{10.3847/2041-8213/ab3d30}

\bibitem[{{Feinstein} {et~al.}(2023){Feinstein}, {Radica}, {Welbanks}, {Murray}, {Ohno}, {Coulombe}, {Espinoza}, {Bean}, {Teske}, {Benneke}, {Line}, {Rustamkulov}, {Saba}, {Tsiaras}, {Barstow}, {Fortney}, {Gao}, {Knutson}, {MacDonald}, {Mikal-Evans}, {Rackham}, {Taylor}, {Parmentier}, {Batalha}, {Berta-Thompson}, {Carter}, {Changeat}, {dos Santos}, {Gibson}, {Goyal}, {Kreidberg}, {L{\'o}pez-Morales}, {Lothringer}, {Miguel}, {Molaverdikhani}, {Moran}, {Morello}, {Mukherjee}, {Sing}, {Stevenson}, {Wakeford}, {Ahrer}, {Alam}, {Alderson}, {Allen}, {Batalha}, {Bell}, {Blecic}, {Brande}, {Caceres}, {Casewell}, {Chubb}, {Crossfield}, {Crouzet}, {Cubillos}, {Decin}, {D{\'e}sert}, {Harrington}, {Heng}, {Henning}, {Iro}, {Kempton}, {Kendrew}, {Kirk}, {Krick}, {Lagage}, {Lendl}, {Mancini}, {Mansfield}, {May}, {Mayne}, {Nikolov}, {Palle}, {Petit dit de la Roche}, {Piaulet}, {Powell}, {Redfield}, {Rogers}, {Roman}, {Roy}, {Nixon}, {Schlawin}, {Tan}, {Tremblin}, {Turner}, {Venot}, {Waalkes}, {Wheatley}, \&
  {Zhang}}]{Feinstein-2023}
{Feinstein}, A.~D., {Radica}, M., {Welbanks}, L., {et~al.} 2023, \nat, 614, 670, \dodoi{10.1038/s41586-022-05674-1}

\bibitem[{Fortney {et~al.}(2020)Fortney, Visscher, Marley, Hood, Line, Thorngren, Freedman, \& Lupu}]{Fortney_2020}
Fortney, J.~J., Visscher, C., Marley, M.~S., {et~al.} 2020, The Astronomical Journal, 160, 288, \dodoi{10.3847/1538-3881/abc5bd}

\bibitem[{Fulton \& Petigura(2018)}]{Fulton_2018}
Fulton, B.~J., \& Petigura, E.~A. 2018, The Astronomical Journal, 156, 264, \dodoi{10.3847/1538-3881/aae828}

\bibitem[{Fulton {et~al.}(2017)Fulton, Petigura, Howard, Isaacson, Marcy, Cargile, Hebb, Weiss, Johnson, Morton, Sinukoff, Crossfield, \& Hirsch}]{fulton_california-kepler_2017}
Fulton, B.~J., Petigura, E.~A., Howard, A.~W., {et~al.} 2017, The Astronomical Journal, 154, 109, \dodoi{10.3847/1538-3881/aa80eb}

\bibitem[{{Gao} {et~al.}(2016){Gao}, {Allen}, {Green}, \& {West}}]{Gao_2016}
{Gao}, C.~W., {Allen}, J.~W., {Green}, W.~H., \& {West}, R.~H. 2016, Comput. Phys. Commun., 203, 212, \dodoi{10.1016/j.cpc.2016.02.013}

\bibitem[{{Graham} {et~al.}(2021){Graham}, {Lichtenberg}, {Boukrouche}, \& {Pierrehumbert}}]{graham2021multispecies}
{Graham}, R.~J., {Lichtenberg}, T., {Boukrouche}, R., \& {Pierrehumbert}, R.~T. 2021, \psj, 2, 207, \dodoi{10.3847/PSJ/ac214c}

\bibitem[{G{\"u}nther {et~al.}(2019)G{\"u}nther, Pozuelos, Dittmann, Dragomir, Kane, Daylan, Feinstein, Huang, Morton, Bonfanti, {et~al.}}]{gunther-2019}
G{\"u}nther, M.~N., Pozuelos, F.~J., Dittmann, J.~A., {et~al.} 2019, Nature Astronomy, 3, 1099

\bibitem[{Hardegree-Ullman {et~al.}(2019)Hardegree-Ullman, Cushing, Muirhead, \& Christiansen}]{Hardegree-Ullman_2019}
Hardegree-Ullman, K.~K., Cushing, M.~C., Muirhead, P.~S., \& Christiansen, J.~L. 2019, The Astronomical Journal, 158, 75, \dodoi{10.3847/1538-3881/ab21d2}

\bibitem[{Heng \& Marley(2018)}]{heng2018radiative}
Heng, K., \& Marley, M.~S. 2018, Radiative Transfer for Exoplanet Atmospheres (Cham: Springer International Publishing), 2137--2152, \dodoi{10.1007/978-3-319-55333-7_102}

\bibitem[{Hu(2019)}]{Hu_2019}
Hu, R. 2019, \apj, 887, 166, \dodoi{10.3847/1538-4357/ab58c7}

\bibitem[{Hu(2021)}]{Hu_2021}
---. 2021, \apj, 921, 27, \dodoi{10.3847/1538-4357/ac1789}

\bibitem[{Hu {et~al.}(2021)Hu, Damiano, Scheucher, Kite, Seager, \& Rauer}]{Hu_2021b}
Hu, R., Damiano, M., Scheucher, M., {et~al.} 2021, \apjl, 921, L8, \dodoi{10.3847/2041-8213/ac1f92}

\bibitem[{Hu \& Seager(2014)}]{Hu_2014}
Hu, R., \& Seager, S. 2014, \apj, 784, 63, \dodoi{10.1088/0004-637X/784/1/63}

\bibitem[{Hu {et~al.}(2012)Hu, Seager, \& Bains}]{Hu_2012}
Hu, R., Seager, S., \& Bains, W. 2012, \apj, 761, \dodoi{10.1088/0004-637X/761/2/166}

\bibitem[{Hu {et~al.}(2013)Hu, Seager, \& Bains}]{Hu_2013}
---. 2013, \apj, 769, 6, \dodoi{10.1088/0004-637X/769/1/6}

\bibitem[{Innes {et~al.}(2023)Innes, Tsai, \& Pierrehumbert}]{Innes-2023}
Innes, H., Tsai, S.-M., \& Pierrehumbert, R.~T. 2023, \apj, 953, 168, \dodoi{10.3847/1538-4357/ace346}

\bibitem[{Johansen \& Lambrechts(2017)}]{Johansen-2017}
Johansen, A., \& Lambrechts, M. 2017, Annual Review of Earth and Planetary Sciences, 45, 359, \dodoi{https://doi.org/10.1146/annurev-earth-063016-020226}

\bibitem[{Johnson {et~al.}(2022)Johnson, Dong, Grinberg~Dana, Chung, Farina, Gillis, Liu, Yee, Blondal, Mazeau, Grambow, Payne, Spiekermann, Pang, Goldsmith, West, \& Green}]{RMG-database}
Johnson, M.~S., Dong, X., Grinberg~Dana, A., {et~al.} 2022, Journal of Chemical Information and Modeling, 62, 4906, \dodoi{10.1021/acs.jcim.2c00965}

\bibitem[{Kempton {et~al.}(2023)Kempton, Lessard, Malik, Rogers, Futrowsky, Ih, Marounina, \& Muñoz-Romero}]{Kempton_2023}
Kempton, E. M.-R., Lessard, M., Malik, M., {et~al.} 2023, The Astrophysical Journal, 953, 57, \dodoi{10.3847/1538-4357/ace10d}

\bibitem[{{Lambrechts, M.} {et~al.}(2014){Lambrechts, M.}, {Johansen, A.}, \& {Morbidelli, A.}}]{Lambrechts-2014}
{Lambrechts, M.}, {Johansen, A.}, \& {Morbidelli, A.} 2014, A\&A, 572, A35, \dodoi{10.1051/0004-6361/201423814}

\bibitem[{Leconte {et~al.}(2017)Leconte, , Selsis, Hersant, \& Guillot}]{Leconte-2017}
Leconte, J., , Selsis, F., Hersant, F., \& Guillot, T. 2017, \aap, 598, A98, \dodoi{10.1051/0004-6361/201629140}

\bibitem[{Leconte {et~al.}(2024)Leconte, Spiga, Clément, Guerlet, Selsis, Milcareck, Cavalié, Moreno, Lellouch, Óscar Carrión-González, Charnay, \& Lefèvre}]{Leconte-2024}
Leconte, J., Spiga, A., Clément, N., {et~al.} 2024, \aap, 686, A131, \dodoi{10.1051/0004-6361/202348928}

\bibitem[{Line {et~al.}(2011)Line, Vasisht, Chen, Angerhausen, \& Yung}]{Line_2011}
Line, M.~R., Vasisht, G., Chen, P., Angerhausen, D., \& Yung, Y.~L. 2011, The Astrophysical Journal, 738, 32, \dodoi{10.1088/0004-637X/738/1/32}

\bibitem[{Liu {et~al.}(2021)Liu, Grinberg~Dana, Johnson, Goldman, Jocher, Payne, Grambow, Han, Yee, Mazeau, Blondal, West, Goldsmith, \& Green}]{rmg-v3}
Liu, M., Grinberg~Dana, A., Johnson, M.~S., {et~al.} 2021, Journal of Chemical Information and Modeling, 61, 2686, \dodoi{10.1021/acs.jcim.0c01480}

\bibitem[{Lodders(2020)}]{Lodders-2020}
Lodders, K. 2020, Solar elemental abundances (Oxford Research Encyclopedia of Planetary Science), \dodoi{10.1093/acrefore/9780190647926.013.145}

\bibitem[{Lodders {et~al.}(2009)Lodders, Palme, \& Gail}]{Lodders-2009}
Lodders, K., Palme, H., \& Gail, H.-P. 2009, Solar system, 712

\bibitem[{Loyd {et~al.}(2016)Loyd, France, Youngblood, Schneider, Brown, Hu, Linsky, Froning, Redfield, Rugheimer, \& Tian}]{Loyd_2016}
Loyd, R. O.~P., France, K., Youngblood, A., {et~al.} 2016, \apj, 824, 102, \dodoi{10.3847/0004-637X/824/2/102}

\bibitem[{Luque \& Pallé(2022)}]{Luque_2022}
Luque, R., \& Pallé, E. 2022, Science, 377, 1211, \dodoi{10.1126/science.abl7164}

\bibitem[{Madhusudhan {et~al.}(2021)Madhusudhan, Piette, \& Constantinou}]{Madhusudhan-2021}
Madhusudhan, N., Piette, A. A.~A., \& Constantinou, S. 2021, \apj, 918, 1, \dodoi{10.3847/1538-4357/abfd9c}

\bibitem[{Madhusudhan {et~al.}(2023)Madhusudhan, Sarkar, Constantinou, Holmberg, Piette, \& Moses}]{Madhusudhan-2023}
Madhusudhan, N., Sarkar, S., Constantinou, S., {et~al.} 2023, ApJL, 956, L13, \dodoi{10.3847/2041-8213/acf577}

\bibitem[{Miller-Ricci {et~al.}(2008)Miller-Ricci, Seager, \& Sasselov}]{Miller-Ricci_2009}
Miller-Ricci, E., Seager, S., \& Sasselov, D. 2008, The Astrophysical Journal, 690, 1056, \dodoi{10.1088/0004-637X/690/2/1056}

\bibitem[{Montet {et~al.}(2015)Montet, Morton, Foreman-Mackey, Johnson, Hogg, Bowler, Latham, Bieryla, \& Mann}]{Montet_2015}
Montet, B.~T., Morton, T.~D., Foreman-Mackey, D., {et~al.} 2015, The Astrophysical Journal, 809, 25, \dodoi{10.1088/0004-637X/809/1/25}

\bibitem[{Morbidelli {et~al.}(2015)Morbidelli, Lambrechts, Jacobson, \& Bitsch}]{Morbidelli-2015}
Morbidelli, A., Lambrechts, M., Jacobson, S., \& Bitsch, B. 2015, Icarus, 258, 418, \dodoi{https://doi.org/10.1016/j.icarus.2015.06.003}

\bibitem[{Morley {et~al.}(2015)Morley, Fortney, Marley, Zahnle, Line, Kempton, Lewis, \& Cahoy}]{Morley_2015}
Morley, C.~V., Fortney, J.~J., Marley, M.~S., {et~al.} 2015, The Astrophysical Journal, 815, 110, \dodoi{10.1088/0004-637X/815/2/110}

\bibitem[{{Moses} {et~al.}(2011){Moses}, {Visscher}, {Fortney}, {Showman}, {Lewis}, {Griffith}, {Klippenstein}, {Shabram}, {Friedson}, {Marley}, \& {Freedman}}]{Moses_2011}
{Moses}, J.~I., {Visscher}, C., {Fortney}, J.~J., {et~al.} 2011, \apj, 737, \dodoi{https://doi.org/10.1088/0004-637X/737/1/15}

\bibitem[{Moses {et~al.}(2013)Moses, Line, Visscher, Richardson, Nettelmann, Fortney, Barman, Stevenson, \& Madhusudhan}]{Moses_2013}
Moses, J.~I., Line, M.~R., Visscher, C., {et~al.} 2013, The Astrophysical Journal, 777, 34, \dodoi{10.1088/0004-637X/777/1/34}

\bibitem[{Moses {et~al.}(2016)Moses, Marley, Zahnle, Line, Fortney, Barman, Visscher, Lewis, \& Wolff}]{Moses_2016}
Moses, J.~I., Marley, M.~S., Zahnle, K., {et~al.} 2016, \apj, 829, 66, \dodoi{10.3847/0004-637X/829/2/66}

\bibitem[{NASA's Exoplanet Archive(2024)}]{NASA_exo_archive}
NASA's Exoplanet Archive. 2024, NASA's Exoplanet Archive.
\newblock \url{https://exoplanets.nasa.gov/}

\bibitem[{Oya {et~al.}(1994)Oya, Shiina, Tsuchiya, \& Matsui}]{Oya-1994}
Oya, M., Shiina, H., Tsuchiya, K., \& Matsui, H. 1994, Bulletin of the Chemical Society of Japan, 67, 2311, \dodoi{10.1246/bcsj.67.2311}

\bibitem[{Peterson {et~al.}(2023)Peterson, Benneke, Collins, Piaulet, Crossfield, Ali-Dib, Christiansen, Gagn{\'e}, Faherty, Kite, {et~al.}}]{peterson-2023}
Peterson, M.~S., Benneke, B., Collins, K., {et~al.} 2023, Nature, 617, 701, \dodoi{https://doi.org/10.1038/s41586-023-05934-8}

\bibitem[{Pierrehumbert(2023)}]{Pierrehumbert-2023}
Pierrehumbert, R.~T. 2023, The Astrophysical Journal, 944, 20, \dodoi{10.3847/1538-4357/acafdf}

\bibitem[{Powell {et~al.}(2024)Powell, Feinstein, Lee, Zhang, Tsai, Taylor, Kirk, Bell, Barstow, Gao, {et~al.}}]{Powell_2024}
Powell, D., Feinstein, A.~D., Lee, E.~K., {et~al.} 2024, Nature, 626, 979, \dodoi{10.1038/s41586-024-07040-9}

\bibitem[{Ranjan {et~al.}(2020)Ranjan, Schwieterman, Harman, Fateev, Sousa-Silva, Seager, \& Hu}]{Ranjan_2020}
Ranjan, S., Schwieterman, E.~W., Harman, C., {et~al.} 2020, \apj, 896, 148, \dodoi{10.3847/1538-4357/ab9363}

\bibitem[{RMG(2023)}]{RMG-developers}
RMG, v. 2023, Developers of Reaction Mechanism Generator and associated software, \url{https://github.com/ReactionMechanismGenerator}

\bibitem[{Rogers \& Seager(2010)}]{Rogers_2010}
Rogers, L.~A., \& Seager, S. 2010, The Astrophysical Journal, 716, 1208, \dodoi{10.1088/0004-637X/716/2/1208}

\bibitem[{{Rustamkulov} {et~al.}(2023){Rustamkulov}, {Sing}, {Mukherjee}, {May}, {Kirk}, {Schlawin}, {Line}, {Piaulet}, {Carter}, {Batalha}, {Goyal}, {L{\'o}pez-Morales}, {Lothringer}, {MacDonald}, {Moran}, {Stevenson}, {Wakeford}, {Espinoza}, {Bean}, {Batalha}, {Benneke}, {Berta-Thompson}, {Crossfield}, {Gao}, {Kreidberg}, {Powell}, {Cubillos}, {Gibson}, {Leconte}, {Molaverdikhani}, {Nikolov}, {Parmentier}, {Roy}, {Taylor}, {Turner}, {Wheatley}, {Aggarwal}, {Ahrer}, {Alam}, {Alderson}, {Allen}, {Banerjee}, {Barat}, {Barrado}, {Barstow}, {Bell}, {Blecic}, {Brande}, {Casewell}, {Changeat}, {Chubb}, {Crouzet}, {Daylan}, {Decin}, {D{\'e}sert}, {Mikal-Evans}, {Feinstein}, {Flagg}, {Fortney}, {Harrington}, {Heng}, {Hong}, {Hu}, {Iro}, {Kataria}, {Kempton}, {Krick}, {Lendl}, {Lillo-Box}, {Louca}, {Lustig-Yaeger}, {Mancini}, {Mansfield}, {Mayne}, {Miguel}, {Morello}, {Ohno}, {Palle}, {Petit dit de la Roche}, {Rackham}, {Radica}, {Ramos-Rosado}, {Redfield}, {Rogers}, {Shkolnik}, {Southworth}, {Teske}, {Tremblin},
  {Tucker}, {Venot}, {Waalkes}, {Welbanks}, {Zhang}, \& {Zieba}}]{Rustamkulov-2023}
{Rustamkulov}, Z., {Sing}, D.~K., {Mukherjee}, S., {et~al.} 2023, \nat, 614, 659, \dodoi{10.1038/s41586-022-05677-y}

\bibitem[{Scheucher \& Hu(2024)}]{Scheucher-2024}
Scheucher, M., \& Hu, R. 2024, in prep

\bibitem[{Scheucher {et~al.}(2020)Scheucher, Wunderlich, Grenfell, Godolt, Schreier, Kappel, Haus, Herbst, \& Rauer}]{Scheucher_2020}
Scheucher, M., Wunderlich, F., Grenfell, J.~L., {et~al.} 2020, The Astrophysical Journal, 898, 44, \dodoi{10.3847/1538-4357/ab9084}

\bibitem[{Shorttle {et~al.}(2024)Shorttle, Jordan, Nicholls, Lichtenberg, \& Bower}]{Shorttle_2024}
Shorttle, O., Jordan, S., Nicholls, H., Lichtenberg, T., \& Bower, D.~J. 2024, The Astrophysical Journal Letters, 962, L8, \dodoi{10.3847/2041-8213/ad206e}

\bibitem[{Sing {et~al.}(2024)Sing, Rustamkulov, Thorngren, Barstow, Tremblin, de~Oliveira, Beck, Birkmann, Challener, Crouzet, {et~al.}}]{sing_2024}
Sing, D.~K., Rustamkulov, Z., Thorngren, D.~P., {et~al.} 2024, Nature, 630, 831, \dodoi{https://doi.org/10.1038/s41586-024-07395-z}

\bibitem[{Thorngren \& Fortney(2019)}]{Thorngren_2019}
Thorngren, D., \& Fortney, J.~J. 2019, The Astrophysical Journal Letters, 874, L31, \dodoi{10.3847/2041-8213/ab1137}

\bibitem[{Thorngren {et~al.}(2016)Thorngren, Fortney, Murray-Clay, \& Lopez}]{Thorngren_2016}
Thorngren, D.~P., Fortney, J.~J., Murray-Clay, R.~A., \& Lopez, E.~D. 2016, The Astrophysical Journal, 831, 64, \dodoi{10.3847/0004-637X/831/1/64}

\bibitem[{Tsai {et~al.}(2018)Tsai, Kitzmann, Lyons, Mendonça, Grimm, \& Heng}]{Tsai_2018}
Tsai, S.-M., Kitzmann, D., Lyons, J.~R., {et~al.} 2018, \apj, 862, 31, \dodoi{10.3847/1538-4357/aac834}

\bibitem[{Tsai {et~al.}(2017)Tsai, Lyons, Grosheintz, Rimmer, Kitzmann, \& Heng}]{Tsai_2017}
Tsai, S.-M., Lyons, J.~R., Grosheintz, L., {et~al.} 2017, Astrophys. J. Suppl. Ser., 228, \dodoi{10.3847/1538-4365/228/2/20}

\bibitem[{Tsai {et~al.}(2021)Tsai, Malik, Kitzmann, Lyons, Fateev, Lee, \& Heng}]{Tsai_2021}
Tsai, S.-M., Malik, M., Kitzmann, D., {et~al.} 2021, \apj, 923, 264, \dodoi{10.3847/1538-4357/ac29bc}

\bibitem[{Tsai {et~al.}(2023)Tsai, Lee, Powell, Gao, Zhang, Moses, Hébrard, Venot, Parmentier, Jordan, Hu, Alam, Alderson, Batalha, Bean, Benneke, Bierson, Brady, Carone, Carter, Chubb, Inglis, Leconte, Lopez-Morales, Miguel, Molaverdikhani, Rustamkulov, Sing, Stevenson, Wakeford, Yang, Aggarwal, Baeyens, Barat, Borro, Daylan, Fortney, France, Goyal, Grant, Kirk, Kreidberg, Louca, Moran, Mukherjee, Nasedkin, Ohno, Rackham, Redfield, Taylor, Tremblin, Visscher, Wallack, Welbanks, Youngblood, Ahrer, Batalha, Behr, Berta-Thompson, Blecic, Casewell, Crossfield, Crouzet, Cubillos, Decin, Désert, Feinstein, Gibson, Harrington, Heng, Henning, Kempton, Krick, Lagage, Lendl, Line, Lothringer, Mansfield, Mayne, Mikal-Evans, Palle, Schlawin, Shorttle, Wheatley, \& Yurchenko}]{Tsai_2023}
Tsai, S.-M., Lee, E. K.~H., Powell, D., {et~al.} 2023, Nature, 617, 483, \dodoi{10.1038/s41586-023-05902-2}

\bibitem[{{Valencia, D.} {et~al.}(2010){Valencia, D.}, {Ikoma, M.}, {Guillot, T.}, \& {Nettelmann, N.}}]{Valencia_2010}
{Valencia, D.}, {Ikoma, M.}, {Guillot, T.}, \& {Nettelmann, N.} 2010, A\&A, 516, A20, \dodoi{10.1051/0004-6361/200912839}

\bibitem[{Van~Eylen {et~al.}(2018)Van~Eylen, Agentoft, Lundkvist, Kjeldsen, Owen, Fulton, Petigura, \& Snellen}]{Van_Eylen_2018}
Van~Eylen, V., Agentoft, C., Lundkvist, M.~S., {et~al.} 2018, Monthly Notices of the Royal Astronomical Society, 479, 4786, \dodoi{10.1093/mnras/sty1783}

\bibitem[{{Venot} {et~al.}(2012){Venot}, {Hébrard}, {Agundez}, {Dobrijevic}, {Selsis}, {Hersant}, {Iro}, \& {Bounaceur}}]{Venot_2012}
{Venot}, O., {Hébrard}, E., {Agundez}, M., {et~al.} 2012, \aap, 546, 1, \dodoi{10.1051/0004-6361/201219310}

\bibitem[{Venturini {et~al.}(2020)Venturini, Guilera, Haldemann, Ronco, \& Mordasini}]{Venturini_2020}
Venturini, J., Guilera, O.~M., Haldemann, J., Ronco, M.~P., \& Mordasini, C. 2020, \aap, 643, L1, \dodoi{10.1051/0004-6361/202039141}

\bibitem[{Welbanks {et~al.}(2024)Welbanks, Bell, Beatty, Line, Ohno, Fortney, Schlawin, Greene, Rauscher, McGill, {et~al.}}]{welbanks_2024}
Welbanks, L., Bell, T.~J., Beatty, T.~G., {et~al.} 2024, Nature, 630, 836, \dodoi{https://doi.org/10.1038/s41586-024-07514-w}

\bibitem[{Wogan {et~al.}(2024)Wogan, Batalha, Zahnle, Krissansen-Totton, Tsai, \& Hu}]{wogan2024jwst}
Wogan, N.~F., Batalha, N.~E., Zahnle, K., {et~al.} 2024, arXiv preprint arXiv:2401.11082

\bibitem[{Woiki \& Roth(1995)}]{Woiki-1995}
Woiki, D., \& Roth, P. 1995, in Shock Waves @ Marseille II (Berlin, Heidelberg: Springer Berlin Heidelberg), 53--58

\bibitem[{Yang {et~al.}(2023)Yang, Gudipati, Henderson, \& Fleury}]{Yang-2023}
Yang, J., Gudipati, M.~S., Henderson, B.~L., \& Fleury, B. 2023, \apj, 947, 26, \dodoi{10.3847/1538-4357/acbd9b}

\bibitem[{Yang \& Hu(2024)}]{Yang_2024}
Yang, J., \& Hu, R. 2024, The Astrophysical Journal, 966, 189, \dodoi{10.3847/1538-4357/ad35c8}

\bibitem[{Yu {et~al.}(2021)Yu, Moses, Fortney, \& Zhang}]{Yu-2021}
Yu, X., Moses, J.~I., Fortney, J.~J., \& Zhang, X. 2021, \apj, 914, 38, \dodoi{10.3847/1538-4357/abfdc7}

\bibitem[{Zahnle {et~al.}(2016)Zahnle, Marley, Morley, \& Moses}]{Zahnle_2016}
Zahnle, K., Marley, M.~S., Morley, C.~V., \& Moses, J.~I. 2016, \apj, 824, 137, \dodoi{10.3847/0004-637X/824/2/137}

\bibitem[{Zhang \& Showman(2018)}]{Zhang_2018}
Zhang, X., \& Showman, A.~P. 2018, The Astrophysical Journal, 866, 1, \dodoi{10.3847/1538-4357/aada85}

\end{thebibliography}
\bibliographystyle{aasjournal}



\end{document}